\pdfoutput=1
\documentclass[acmlarge]{acmart}

\usepackage{booktabs}   
\usepackage{subcaption} 
\usepackage{comment}
\usepackage{amsmath}
\usepackage{graphicx}
\usepackage{graphics}
\usepackage{adjustbox}
\usepackage{caption}
\usepackage{algorithmicx}
\usepackage{algpseudocode}
\usepackage{algorithm}
\usepackage{amsfonts}
\usepackage{amssymb}
\usepackage{listings}
\usepackage{color}
\usepackage{epstopdf}
\usepackage{syntax}
\usepackage{multicol}
\usepackage{url}
\usepackage{hyperref}
\usepackage{wrapfig}

\makeatletter\if@ACM@journal\makeatother
 \acmJournal{PACMPL}
 \acmVolume{1}
 \acmNumber{1}
 \acmArticle{1}
 \acmYear{2017}
\acmMonth{1}
 \acmDOI{10.1145/nnnnnnn.nnnnnnn}
 \startPage{1}
 \else\makeatother
\acmConference[PL'17]{ACM SIGPLAN Conference on Programming Languages}{January 01--03, 2017}{New York, NY, USA}
\acmYear{2017}
 \acmISBN{978-x-xxxx-xxxx-x/YY/MM}
 \acmDOI{10.1145/nnnnnnn.nnnnnnn}
 \startPage{1}
 \fi

\setcopyright{none}             

\DeclareCaptionType{copyrightbox}
\floatname{algorithm}{Procedure}

\definecolor{mygreen}{rgb}{0,0.6,0}
\definecolor{mygray}{rgb}{0.5,0.5,0.5}
\definecolor{mymauve}{rgb}{0.58,0,0.82}

\begin{document}

\title{Data-Driven Program Completion}         


\author{Yanxin Lu}
\affiliation{
  \institution{Rice University}            
  \streetaddress{MS-132}
  \city{Houston}
  \state{TX}
  \postcode{77005}
  \country{USA}
}
\email{yanxin.lu@rice.edu}          

\author{Swarat Chaudhuri}
\affiliation{
  \institution{Rice University}           
  \city{Houston}
  \state{TX}
  \postcode{77005}
  \country{USA}
}
\email{swarat@rice.edu}         

\author{Chris Jermaine}
\affiliation{
  \institution{Rice University}           
  \city{Houston}
  \state{TX}
  \postcode{77005}
  \country{USA}
}
\email{cmj4@rice.edu}         

\author{David Melski}
\affiliation{
  \institution{Grammatech, Inc.}           
  \city{Ithaca}
  \state{NY}
  \postcode{14850}
  \country{USA}
}
\email{melski@grammatech.com}         


\begin{abstract}
  We introduce {\em program splicing}, a programming methodology that
  aims to automate the commonly used workflow of copying, pasting, and
  modifying code available online. Here, the programmer starts by
  writing a ``draft'' that mixes unfinished code, natural language
  comments, and correctness requirements in the form of test cases or
  API call sequence constraints. A program synthesizer that interacts
  with a large, searchable database of program snippets is used to
  automatically complete the draft into a program that meets the
  requirements. The synthesis process happens in two stages. First,
  the synthesizer identifies a small number of programs in the
  database that are relevant to the synthesis task. Next it uses an
  enumerative search to systematically fill the draft with expressions
  and statements from these relevant programs. The resulting program
  is returned to the programmer, who can modify it and possibly invoke
  additional rounds of synthesis.


  We present an implementation of program splicing for the Java
  programming language. The implementation uses a corpus of over 3.5
  million procedures from an open-source software repository. Our
  evaluation uses the system in a suite of everyday programming tasks,
  and includes a comparison with a state-of-the-art competing approach
  as well as a user study. The results point to the broad scope and
  scalability of program splicing and indicate that the approach can
  significantly boost programmer productivity.

\end{abstract}




\maketitle

\setlength{\columnsep}{15pt}
\section{Introduction}
\label{sec:intro} 

Copying and pasting from existing code is a coding practice that
refuses to die out in spite of much expert
disapproval~\cite{kim2004ethnographic,juergens2009code}. The approach
is vilified for good reason: it is easy to write buggy programs using
blind copy-and-paste. At the same time, the widespread nature of the
practice indicates that programmers often have to write code that
substantially overlaps with existing code, and that they find it
tedious to write this code from scratch.

In spite of its popularity, copying and pasting code is not always
easy. To copy and paste effectively, the programmer has to 
identify a piece of code that is relevant to their work. After pasting
this code, they have to modify it to fit the requirements of their
task and the code that they have already written. Many of the bugs
introduced during copying and pasting come from the low-level, manual
nature of the task.

In this paper, we present a programming methodology, called {\em
  program splicing}, that aims to offer the benefits of copy-and-paste
without some of its pitfalls.  Here, the programmer writes code with
the assistance of a program synthesizer~\cite{Alur2015,lezama06} that is able to query a large,
searchable database of program snippets extracted from online
open-source repositories. Operationally, the programmer starts by
writing a ``draft'' that is a mix of unfinished code and natural
language comments, along with an incomplete correctness requirement,
for example in the form of test cases or API call sequence constraints. The
synthesizer completes the ``holes'' in the draft by instantiating them
with code extracted from the database, such that the resulting program
meets its correctness requirement. The programmer may then further
modify the program and possibly proceed to perform additional rounds
of synthesis.

\begin{wrapfigure}{r}{0.4\textwidth}
  \vspace{-10pt}
  \begin{lstlisting}
/* COMMENT:
 * use sieve of eratosthenes
 * to test primality
 * TEST:
 * __solution__
 * return sieve(1) == false &&
 *        sieve(2) == true &&
 *        sieve(29) == true;
 */
boolean sieve(int num) {
  boolean[] prime=new boolean[N];
  for(int i = ??; i <= num; ++i)
    prime[i] = ??;
  // build a table
  ??;
  return prime[num];}
  \end{lstlisting}
  \vspace{-10pt}
  \caption{Primality Testing: Draft}
  \label{fig:primality-sketch-1}
  \vspace{-10pt}
\end{wrapfigure}

In more detail, our synthesis algorithm operates as follows. First, it
identifies and retrieves from the database a small number of program
snippets that are relevant to the code in the draft.  These search
results are viewed as pieces of knowledge relevant to the synthesis
task at hand, and are used to guide the synthesis
algorithm. Specifically, from each result, the algorithm extracts a
set of {\em codelets}: expressions and statements that are conceivably
related to the synthesis task. Next, it systematically enumerates over
possible instantiations of holes in the draft with codelets, using a
number of heuristics to prune the space of instantiations.


The primary distinction between our synthesis algorithm and existing
search-based approaches to synthesis lies in the use 
of pre-existing code.
A key benefit of such a data-driven approach is that it helps with the
problem of {\em underspecification}. Because synthesis involves the
{\em discovery} of programs, the specification for a synthesis problem
may be incomplete. This means that even if a synthesizer finds a
solution that meets the specification, this solution may in fact be
nonsensical. This problem is especially common in traditional
synthesis tools, which explore a space of candidate programs without
significant human guidance. In contrast, the codelets in our approach
are sourced from pre-existing code that humans wrote when solving
related programming tasks. This means that our search for programs is
biased towards programs that human-readable and likely to follow
common-sense constraints that humans assume.

\begin{wrapfigure}{r}{0.4\textwidth}
  \begin{lstlisting}
boolean sieve(int num) {
  boolean[] prime = new boolean[N];
  // build a table
  for(int i=2; i<=num; i++)
    prime[i]=true;
  for(int i=2; i<=num/2; i++)
    for(int j=2; j<=num/i; j++)
      prime[i*j]=false;
  return prime[num];
}
  \end{lstlisting}
  \vspace{-15pt}
  \caption{Primality Testing: Completed Draft}
  \label{fig:primality-sketch-2}
  \vspace{-10pt}
\end{wrapfigure}

The use of pre-existing code also has a positive effect on
scalability.  Without codelets, the synthesizer would have to
instantiate holes in the draft with expressions built entirely from
scratch. In contrast, in program splicing, the synthesizer searches
the more limited space of ways in which codelets can be ``merged''
with pre-existing code.

We present an implementation of program splicing that uses a corpus of
approximately 3.5 million methods, extracted from the
Sourcerer~\cite{sajnani:icsem2014, Ossher:WCRE12, Bajracharya:SCP14}
source code repository, to perform synthesis of Java programs.  We
evaluate our approach on a suite of Java programming tasks, including
the implementation of scripts useful in everyday computing,
modifications of well-known algorithms, and initial prototypes of
software components such as GUIs, HTML parsers, and HTTP servers. Our
evaluation includes a comparison with $\mu$Scalpel~\cite{Barr2015}, a
state-of-the-art programming system that can ``transplant'' code
across programs, as well
as a user study with 18 participants.  The evaluation shows our system
to outperform $\mu$Scalpel and indicates that it can significantly boost
overall programmer productivity.

Now we summarize the contributions of the paper:
\begin{itemize}
  \item We propose program splicing, a methodology where
    programmers use a program synthesizer that can query
    a large database of existing code, as a more robust proxy for
    copying and pasting code. 
  \item We present an implementation of program splicing for the Java
    language that is driven by a corpus of 3.5 million Java methods.
  \item We present an extensive empirical evaluation of our system on a range of
    everyday programming tasks. The evaluation, which includes a user
    study, shows that our method outperforms a state-of-the-art
    competing approach and increases overall programmer productivity.
\end{itemize}

The rest of the paper is organized as follows. In Section 2, we give
an overview of our method. In Section 3, we formally state our
synthesis problem. Section 4 describes the approach of program
splicing. Section 5 presents our evaluation. Related work is described
in Section 6. We conclude with some discussion in Section 7.

\section{Overview}
\label{sec:overview}


In this section, we describe program splicing from a user's
perspective using a few motivating examples.

\begin{wrapfigure}{r}{0.4\textwidth}
  \vspace{-5pt}
  \begin{lstlisting}
void sieve(boolean[] p) {
  p[1] = false;
  int l = p.length - 1;
  for(int i=2; i<=l; i++)
    p[i]=true;
  for(int i=2; i<=l/2; i++)
    for(int j=2; j<=l/i; j++)
      p[i*j]=false;
}
  \end{lstlisting}
\vspace{-15pt}
\caption{Sieve of Eratosthenes Algorithm}
\label{fig:primality-pdb}
\vspace{-10pt}
\end{wrapfigure}




\subsection{Primality Testing} Consider a programmer who would like to
implement a primality testing function using the Sieve of Eratosthenes
algorithm. The programmer knows that the function must build an array
{\tt prime} of bits, the $i$-th bit being set to true if the number
$i$ is a prime. However, they do not recall in detail how to
initialize the array and the algorithm for populating this array.


In current practice, the programmer would search the web for a Sieve
of Eratosthenes algorithm, copy code from one of the search results,
and modify this code manually. In contrast, in program splicing, they
write a {\em draft} program in a notation inspired by the Sketch
system for program synthesis~\cite{lezama06,solar2009sketching}
(Figure~\ref{fig:primality-sketch-1}).  This draft program declares
the array {\tt prime}; however, in place of the code to fill this
array, simply leaves a {\em hole} represented by a special symbol
``\verb|??|''.  A hole in a program serves as a placeholder for an
external codelets which will be filled in by our system. In this
example, the external snippets will be an Sieve of Eratosthenes
implementation.

\begin{wrapfigure}{r}{0.4\textwidth}
  \vspace{-10pt}
  \begin{lstlisting}
int[][] read_csv(int[][] m,
    int r, int c, String filename) {
  File f = new File(filename);
  Scanner scanner = new Scanner(f);
  for(int i = 0; i < r; ++i) {
    String line=scanner.nextLine();
    String[] fields=line.split(",");
    for(int j = 0; j < c; ++j)
      m[i][j] =
        Integer.parseInt(fields[j]);
  }
  return m;
}
  \end{lstlisting}
  \vspace{-20pt}
  \caption{Reading CSV: Reading a matrix from a CSV file}
  \label{fig:csv-allocation}
  \vspace{-10pt}
\end{wrapfigure}

The user
describes the forms of external code that are relevant to the task
using natural language comments. 
In this example, the comments contain words such as ``sieve'',
``eratosthenes'' and ``primality'' in the ``\verb|COMMENT|'' section
at line 1 suggesting a Sieve of Eratosthenes implementation. The
system will use these words as a hint to search the code
database. This is similar to a web search using text, but in this case
it is done in a programming scenario. Finally, in order to ensure that
the synthesized code is compatible with the code that she has already
written, the programmer needs to provide some correctness
requirements. The requirements for our example are shown in the
``\verb|TEST|'' section at the top of the draft.

Given the draft, our program synthesizer issues a query to a
searchable database of code snippets. The code database then returns a
set of functions relevant to the current programming task, including
at least one Sieve of Eratosthenes implementation (such an
implementation is shown in Figure~\ref{fig:primality-pdb}). The system
now extracts a set of {\em codelets} --- expressions and statements
--- from these functions, and uses a composition of these codelets to
fill in the hole in the draft. The completed draft is showed in
Figure~\ref{fig:primality-sketch-2}.

\begin{figure*}
    \centering
    \begin{subfigure}[b]{0.45\textwidth}
      \begin{lstlisting}
int[][] csvmat(String filename) {
  int[][] matrix = new int[N][N];
  /* COMMENT:
   * Read a matrix from a csv file
   * TEST:
   * String filename = ``matrix.csv'';
   * int [][] matrix = new int[N][N];
   * __solution__
   * return test_matrix(matrix);
   */
  ??
}
      \end{lstlisting}
      \vspace{-20pt}
      \caption{Reading CSV: draft for reading from CSV}
      \label{fig:csv-sketch-1}
    \end{subfigure}
    \begin{subfigure}[b]{0.45\textwidth}
      \begin{lstlisting}
int[][] csvmat(String filename) {
  int[][] matrix = new int[N][N];
  ...
  int[][] mat = new int[N][N];
  /* COMMENT:
   * matrix multiplication
   * TEST:
   * int[][] matrix={{1, 2, 3}, ...};
   * int[][] result={{14, 20, 26}, ...};
   * __solution__
   * return test_equality(matrix, result);
   */
  ??
  return mat;
}
      \end{lstlisting}
      \vspace{-25pt}
      \caption{Reading CSV: draft for matrix multiplication}
      \label{fig:csv-sketch-2}
    \end{subfigure}
    \caption{Reading CSV draft programs}
    \label{fig:sketches}
    \vspace{-15pt}
\end{figure*}

%
%

\subsection{Reading a Matrix from a CSV File} 
Now we show an example where external code snippets are used to
complete a draft with multiple holes, through an interactive
process. Suppose the programmer would like to read a matrix from a
comma-separated values (CSV) file into a 2-dimensional array and then
to square the matrix. This programming task has two major pieces:
reading from the csv file and matrix multiplication. In the beginning,
the programmer focuses on the first task, and accordingly, writes the
draft program shown in Figure~\ref{fig:csv-sketch-1}. In this draft,
the programmer simply declares a 2d-array. Then she leaves a hole as
proxy for the code for reading the matrix from the csv file, and
provides some comments and requirements to guide the instantiation of the
hole. Our system then searches the code database for relevant external
code. For example, such a program is shown in
Figure~\ref{fig:csv-allocation}.  Snippets from this code is then
merged into the existing draft.

\begin{wrapfigure}{r}{0.5\textwidth}
  \vspace{-10pt}
  \begin{lstlisting}
int[][] csvmat(String filename) {
  int[][] matrix = new int[N][N];
  File f = new File(filename);
  Scanner scanner = new Scanner(f);
  for(int i = 0; i < N; ++i) {
    String line = scanner.nextLine();
    String[] fields = line.split(",");
    for(int j = 0; j < N; ++j)
      matrix[i][j]=Integer.parseInt(fields[j]);
  }
  int[][] mat = new int[N][N];
  for(int i = 0; i < N; ++i)
    for(int j = 0; j < N; ++j) {
      int s = 0;
      for(int k = 0; k < N; ++k) 
        s += matrix[i][k] * matrix[k][j];
      mat[i][j] = s;
    }
  return mat;}
  \end{lstlisting}
  \vspace{-20pt}
\caption{Reading CSV: Complete Program}
\label{fig:csv-sketch-3}
\end{wrapfigure}

After getting the code that reads a matrix from a csv file, the user
now focuses on the second part of the task, which is matrix
multiplication. They extend the previous code into a new draft, which
has a hole for the matrix multiplication code, some comments and
requirements. This draft is shown in Figure~\ref{fig:csv-sketch-2}. Our
system now searches the code database for codelets that does matrix
multiplication and merges these codelets into the existing code, while
ensuring that all requirements are met. The complete program resulting
from this process is shown in Figure~\ref{fig:csv-sketch-3}.

As shown in the example, our system can be used in an iterative and
interactive manner. A programmer can start writing code as usual, and
brings in external resources from the web into the existing codebase
as needed. In this respect our approach is similar to
copy-and-paste. The difference is that our system automates the
process of finding and modifying relevant code, and guarantees a
certain level of reliability by ensuring that the output program
meets all its requirements.

\subsection{Face Detection using OpenCV} 

In previous examples, we rely on input-output tests to verify the
correctness of a solution. Now we consider the use of program splicing
in the implementation of {\em face detection}, a computer vision task
in which input-output tests are hard to specify, requiring the use of
an alternative form for correctness requirement. Specifically, the
requirements that we use are constraints on sequences of API calls
that a program makes, given in the form of a finite automaton. 

\begin{wrapfigure}{r}{0.45\textwidth}
  \begin{lstlisting}
/* COMMENT:
 * Doing face detection using OpenCV
 * TEST:
 * API_cons("FaceDetectionTest.java");
 * __solution__
 * run();
 * test(_has_detector_  &&
 *      _has_image_  &&
 *      _has_detection_ &&
 *      _image_written_);
 */
public void run() {
  String input_img = "lena.jpg";
  String output_img = "faceDetection.png";
  CascadeClassifier faceDetector = new CascadeClassifier(??);
  ??
}
  \end{lstlisting}
  \vspace{-40pt}
  \caption{Face Detection: Draft Program}
  \label{fig:face-sketch-1}
  \vspace{-20pt}
\end{wrapfigure}

\begin{wrapfigure}{r}{0.65\textwidth}
  \vspace{-10pt}
  \includegraphics[scale=0.3]{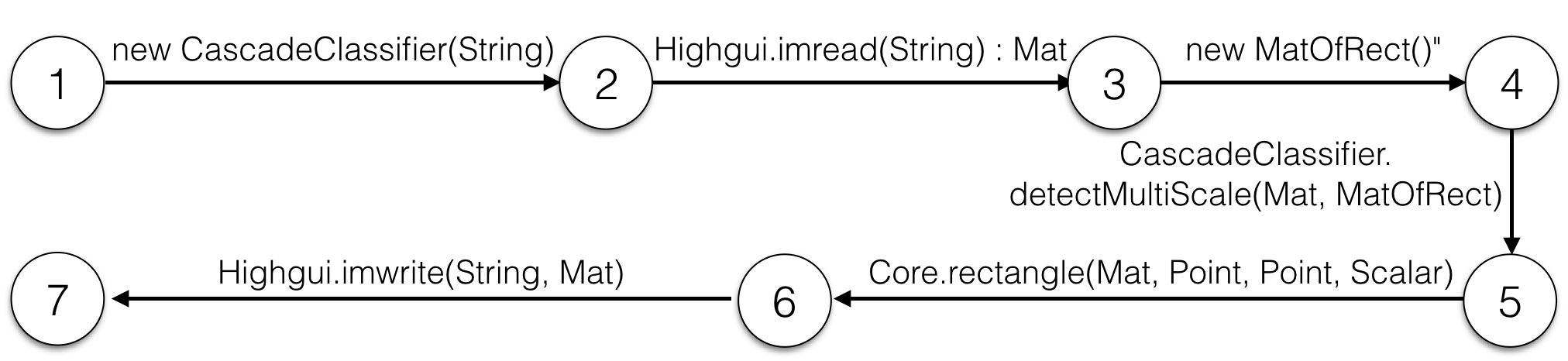}
  \caption{API call sequence constraint for face detection}
  \label{fig:face-spec}
  \vspace{-10pt}
\end{wrapfigure}

Figure~\ref{fig:face-sketch-1} shows a draft program for this task. In
this example, a user wants to use a \verb|CascadeClassifier| object from
OpenCV to detect faces from an input image called \verb|lena.jpg|. The
output image named \verb|faceDetection.png| should have the same
picture with a rectangle drawn above the faces.
The API call constraint for the task is shown in
Figure~\ref{fig:face-spec}. This requirement describes a sequence of
object creation and API invocation actions performed during face
detection. While the requirement is more low-level than unit tests, we
note that it frees users from specifying small details such as what
configuration file to be used, the color for drawing rectangles on
faces and the order of specifying the four corners of rectangles. Our
synthesizer uses this requirement to filter out many of the candidate
programs that it considers during synthesis. Only a few solutions
satisfy the requirement, and the user could easily pick the correct
one shown in Figure~\ref{fig:face-complete}.

\begin{figure}
  \begin{lstlisting}
public void run() {
  String input_image = "lena.png";
  String filename = "faceDetection.png";
  // Create a face detector from the cascade file in the resources
  // directory.
  CascadeClassifier faceDetector =
    new CascadeClassifier(getClass().getResource("lbpcascade_frontalface.xml").getPath());
  Mat image = Highgui.imread(getClass().getResource(input_image).getPath());
  // Detect faces in the image.
  // MatOfRect is a special container class for Rect.
  MatOfRect faceDetections = new MatOfRect();
  faceDetector.detectMultiScale(image, faceDetections);
  // Draw a bounding box around each face.
  for (Rect rect : faceDetections.toArray()) {
    Core.rectangle(image, new Point(rect.x, rect.y), new Point(rect.x + rect.width, rect.y + rect.height), new Scalar(0, 255, 0));
  }
  // Save the visualized detection.
  Highgui.imwrite(filename, image);
  }}
  \end{lstlisting}
  \vspace{-20pt}
  \caption{Face Detection: Complete Program}
  \label{fig:face-complete}
  \vspace{-15pt}
\end{figure}



\section{Problem formulation}
\label{sec:problem}

In this section, we formulate the program synthesis problem that is at
the heart of the program splicing methodology.

\paragraph{Language Definition} 
As mentioned earlier, a draft program in our setting consists of
incomplete code and a set of natural language comments. We start
by spelling out the language of code permitted in our drafts.

Our approach accepts code in a subset $\mathcal{L}$ of Java,
abstractly represented by the following grammar. In summary, the
grammar permits standard imperative expressions and statements over
base and array types, as well as a symbol \verb|??| representing
holes.

\begin{grammar}
  <expr> ::= \emph{id} | \emph{c} | <expr> \emph{binop} <expr> | \emph{unaryop} <expr> | f(<expr>, \dots, <expr>) | \emph{id} := <expr> | ??

  <stmt> ::= \textbf{let} \emph{id} = <expr> | \textbf{if} <expr> <stmt> <stmt> | \textbf{while} <expr> <stmt> | <stmt> ; <stmt> | ??
  
  <program> ::= \emph{id} (<expr>, \dots, <expr>) <stmt>
\end{grammar}

In this grammar, $c$ represents a constant, $id$ represents an
identifier, $f$ represents external functions (API calls), and
$\mathit{binop}$ and $\mathit{unaryop}$ respectively represent binary
and unary operators. We assume that a standard type system is used to
assign types to expressions and statements in this grammar. The actual
language handled by our implementation goes somewhat beyond this
grammar, permitting arrays, objects, data structure definitions, a
limited form of recursion, and syntactic sugar such as for-loops.

The special symbol \verb|??| in the grammar represents two kinds of
holes. \emph{Expression holes} is a placeholder for a missing
expression. A \emph{statement hole} is a placeholder for a missing
statement. 

The semantics of a program with holes can be defined as a set of
complete (hole-free) programs obtained by instantiating the holes with
expressions and statements. The semantics of a complete program is
defined in the standard way. We skip the formal definitions of these semantics
for brevity.


%
%
%
%

\paragraph{Requirement} Aside from a draft, an input to a program
splicing problem includes a {\em requirement}. This requirement is not
expected to be a full correctness specification. Specifically, our
implementation permits two classes of requirements: input-output
tests, and finite automata that constrain the sequences of API calls
that a program can make. We assume a procedure to conservatively check
whether a given complete program satisfies a given set of
requirements.  For requirements that are input-output tests, this
procedure simply evaluates the program on the tests. The procedure for
automaton constraints is based on a standard, sound program analysis.

\paragraph{Synthesis Problem} Let $P_s \in \mathcal{L}$ be a draft
program with one or more holes. Let $DB \subseteq \mathcal{L}$ be a
database containing programs with {\em no} holes. Our objective is to
use the programs from $DB$ to complete holes in $P_s$. Specifically,
we use the expressions (similarly, statements) from $DB$ to
complete the expression holes (similarly, statement holes) in
$P_s$. Naturally, such an instantiation of the holes can be performed
in many different ways. Our goal is to do this instantiation such that
the resulting program passes the requirement.


More precisely, consider the set $\mathcal{C}$ of all {\em codelets}
--- subexpressions and statements --- that appear in programs from
$DB$. Let $\mathcal{P}$ be the set of complete programs obtained by
instantiating the holes of $P_s$ by appropriately typed codelets in
$\mathcal{C}$.  Let
$U: \mathcal{L} \rightarrow \{\mathit{True}, \mathit{False}\}$ be a
function that maps a complete program in $\mathcal{L}$ to a boolean
value indicating whether the input program passes the requirement 
accompanying $P_s$.  Our problem is to find a program
$P^*_c \in \mathcal{P}$ such that $U(P^*_c) = \mathit{True}$.

\section{Method}
\label{sec:method}

In this section, we present our solution to the synthesis problem
presented in the previous section. 

Our synthesis problem has two key 
subproblems: {\em code search} and {\em hole substitution}. 
\begin{itemize}
\item {\em Code Search:} Given a program $P_s \in \mathcal{L}$, search
  a large corpus containing thousands of programs for a set of
  relevant programs such that the retrieved programs contain the
  codelets we want to synthesize. The desired properties of the code
  search technique should be high precision and high efficiency. Here,
  we define precision as the number of retrieved programs that have
  the codelet we want to synthesize, and we define efficiency as
  the runtime required for each search. In summary, we need to fetch a
  set of programs which contain the exact codelet we want within a
  short period of time.

\item {\em Hole Substitution:} Given multiple database programs $S_d$,
  we would like to search for the correct codelets to substitute the
  hole. Multiple programs combined consist of a large number of
  codelets. The key challenge here is to prune the search space such
  that we can efficiently get the exact codelet we want and ensure the
  codelets we want will not be dropped by our heuristics.
\end{itemize}

In general, solving the synthesis problem requires us to find a sweet spot between
expressiveness and efficiency. In traditional program synthesis, an
algorithm with high expressiveness is more capable of generating
various kinds of code, but usually requires more time to search. In
contrast, an efficient algorithm tends to generate a very limited
amount of code, and sometimes the search space fails to cover the
solution.

In our case, expressiveness corresponds to the number of codelets from
the database programs we consider during synthesis and the efficiency
corresponds to the time the synthesis task takes or the number of
incorrect programs we filter in the end. Ideally, we want a
sufficiently expressive and efficient synthesis algorithm such that it
can complete any draft program within a short amount of time. In this
section, we start explain the detail of each component of our
method. We first discuss how our code search method gives us
sufficient expressiveness and how efficiency is achieved with the
synthesis algorithm without sacrificing too much expressiveness.

\subsection{Searching for programs}
\label{subsection:search}

\newcommand{\vknote}[1]{\textbf{[FIXME: [#1]]}}

In this section, we describe the code search techniques employed to
query a large database of programs effectively. This is the first step
in our workflow: to find candidate functionality from the program
database to complete the draft program. Given a draft program
with a hole, relevant code based on the context (such as comments,
function signature, parameter names) around the hole are returned by
the code search.

An important goal of the code search component is to have quick
response when searching large amounts of code to ensure efficiency of
our synthesis algorithm. To accomplish this, various code features are
extracted from a large corpus of open source code. These code
features---along with the corresponding source code---are stored in a
program database. The program database is a scalable object-store
database that allows for fast similarity-based queries.

A query issued to the program database includes code features
extracted from the draft program, along with associated weights
indicating the relative importance of the code features. The program
database computes the k-nearest neighboring corpus elements to the
query, using the code features stored, associated weights, and
similarity metrics defined on each code feature. The result of the
query is presented as a ranked list of source code corresponding to
the k-nearest neighbors.

Expressiveness can be easily guaranteed, since we control the number
of neighbors we consider. We can increase $k$ until we have enough
programs which contains the codelet we want to synthesize. Notice that
the more programs we retrieve, the larger the search space is, and
thus synthesis will require much more time. Eventually, we need to
target at a small $k$ and ensure we have the desired programs.

Below we describe the features extracted and the associated similarity
metrics.

\textbf{Natural language terms.}
For this feature, we extract the function name, comments, local
variable names, and parameter names of a function. Such extracted
natural language (NL) terms are then subjected to a series of standard
NL pre-processing steps, such as splitting words with underscores or
camel-case, removing stop words (including typical English stop words,
and those specialized for Java code), stemming, lemmatization, and
removing single character strings.

Additionally, we use a greedy algorithm~\cite{feild2006empirical} for
splitting terms into multiple words, based on dictionary lookup. This
is to handle the case where programmers combine multiple words,
without separating the words with underscores or camel-case, when
naming functions and variables.

After NL pre-processing, we compute a tf-idf (term frequency-inverse
document frequency) score for each NL term. Each function is
considered as a document, and the tf-idf is computed per project. We
give the function name term an inflated score ($5\times$ more than
other terms) because it often provides significant information about a
function's purpose.

The similarity between two functions is measured by taking the
cosine-similarity of their NL terms, together with their tf-idf
values. Below is an example of NL terms features for the draft showed
in figure~\ref{fig:primality-sketch-1}.

\begin{verbatim}
"primal":0.10976425998969035, "siev":0.658585559938142, "test":0.10976425998969035, 
"prime":0.658585559938142, ...
\end{verbatim}


\textbf{Names.} Here, we extract all the variable names, the name of the
function, and perform some basic normalization such as splitting camel
case and underscores. The similarity metric used is the Jaccard index
on sets of names.


The similarity search is primarily driven by the natural language term
features, with variable names and function names providing
additional context around the hole in the query code. We give more
weights to natural language term features and less weights to
variable names and function names. The reason is that the most
important hint in the source code is comment, because users are
required to describe the code they want to synthesize. However,
variable names and function names must not be treated as equally
important, because sometimes variable names and function names
might be totally irrelevant to the code they want to synthesize. For
example, users might leave comments saying that they want the code
that reads a matrix from a csv file, but it is totally possible that the
surrounding context is all about matrix calculation.

\subsection{Program completion}
%
%
%


After we have retrieved a set of programs from the program database,
our next step is to complete the draft. For each database program
paired with the given partial program, we spawn a thread to do the
code completion task, parallelizing the process in order to have high
efficiency. A code completion task consists of the following steps:

\subsubsection{Hole substitution}
The first step is to use the codelets from the database program to
substitute the holes in the draft. Procedure~\ref{alg:fill} shows the
algorithm. We start by checking whether there is any hole in the draft
at line 1. If not, we move on to the merging step. Otherwise, we start
injecting codelets into the draft. For each hole, we iterate all the
codelets starting from the smallest one and check whether the
injection is valid using our heuristics at line 6. If so, we then
substitute the hole with the codelet at line 7 and then continue
injecting more codelets by recursively calling itself at line 8 until
we finish filling all the holes. When no more holes exist in the draft
program, we then merge the codelets into the existing codebase, which
is explained in detail in later section. If at some point injecting a
codelet is not successful, we backtrack and try another codelet.

\begin{wrapfigure}{R}{0.38\textwidth}
  \vspace{-20pt}
\begin{minipage}{0.38\textwidth}
\begin{algorithm}[H]
  \caption{fill}
  \begin{algorithmic}[1]
    \Require A draft program, $P_s \in \mathcal{L}$ and a database
    program $P_d \in \mathcal{L}$
    \Ensure A complete program $P_c$

    \If {not $\text{has\_hole}(P_s)$}
      \State \Return $\text{merge}(P_s)$
    \EndIf

    \For {$h \gets \text{next\_hole}(P_s)$}
      \For {$n \gets \text{next\_codelet}(P_d)$}
        \If {valid($P_s, h, n$)}
          \State $P_s' \gets \text{substitute}(P_s, h, n)$
          \State $P_c \gets \text{fill}(P_s', P_d)$
          \If {$P_c \neq null$}
            \State \Return $P_c$
          \EndIf
        \EndIf
      \EndFor
    \EndFor
    \State \Return $null$
  \end{algorithmic}
  \label{alg:fill}
\end{algorithm}
\end{minipage}
\end{wrapfigure}

Our search space is then over a finite set of codelets, giving us
guaranteed termination. However, we would still like to apply some
heuristics to make the synthesis more efficient, because the search
space is still quite large given that we need to have a substantial
amount of database program to guarantee expressiveness. Next, we
discuss our heuristics used in the step of hole substitution.

\textbf{Synthesizing expressions} If we are searching for
substitutions for an expression hole $h$, we can first infer the type
of $h$ using surrounding context. If we try to use an expression
codelet $n$ from a database program, we need to first ensure $h$ and
$n$ are of the same type. Otherwise, we ignore $n$. This heuristic
actually gives us efficiency at no cost of expressiveness.

\begin{wrapfigure}{r}{0.6\textwidth}
  \centering
  \includegraphics[scale=0.35]{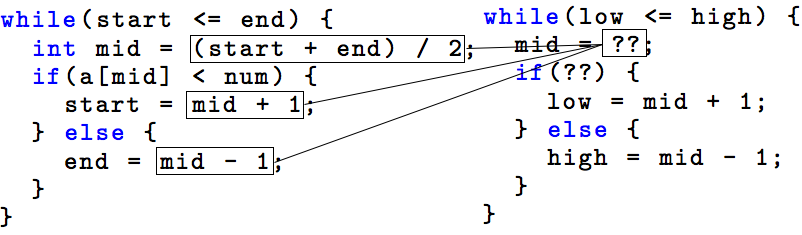}
  \caption{Matching for expression codelet}
  \label{fig:matching}
\end{wrapfigure}

In addition, we can also consider the \emph{roles} of $h$ and $n$. The
intuition is that we only consider the codelet that serves as the same
role by looking at the parent of $n$ and the parent of $h$ in the
parse tree. If the parents of $n$ and $h$ are not of the same kind,
then we discard $n$ and look for another
codelet. Figure~\ref{fig:matching} illustrates the idea. If we are
looking for a codelet to replace a hole representing the \verb|rval|
inside an assignment statement, our target codelets are more likely to
be the \verb|rval| of other assignment statements. We can then just
consider those codelets as substitutions and ignore other
codelets. The same can be applied if we want to synthesize the code
for the guard of a condition.

This heuristic also gives us better efficiency, but some
expressiveness is reduced, because an expression with different role
might still be the desired one. In addition, an algorithm has
different implementations, and therefore it is possible that we might
throw away useful expressions because of different program
structures. However, we can increase the number of database programs
to cover enough variations such that it is more likely to have the
expressions we want. Therefore, it is safe to sacrifice some
expressiveness for efficiency.

\textbf{Synthesizing statements} When we are searching for
substitutions for a statement hole $h$, we need to consider a sequence
of statements from the database program. We first define a sliding window
of various lengths and use that to scan the database program in order
to identify the statement sequence we would like to use to substitute $h$.
We also scan the sequences under loops and conditions. We then
use each codelet to substitute the hole.

We considered using the heuristic similar to role matching for
expression holes, but later we discovered that role matching cannot be
applied when we are synthesizing statement sequences for statement
holes. Typically, the semantic of an expression $e$
includes $e$'s surrounding context and the meaning of $e$ varies if
its surrounding context or role changes. An expression by itself tends
to not have any useful meaning until it is tied to a specific role in
a semantic structure. Therefore surrounding context could be
indicative in selecting a target expression.

However, that is not the case for statement holes. Most statement
sequences tend to serve as a stand-alone functionality and its
semantic is rather complete, and they can appear anywhere regardless
of its surrounding context. Surrounding context in this case does not
provide any useful information and sometimes it becomes even
misleading. Hence, using additional surrounding context tends to
undermine the synthesis algorithm. Consider the following two programs
where we want to generate a codelet under a loop, but the code we want
to synthesize is at the top level of the function. If we were checking
the parents of these two codelets, we would be throwing out the code
we want, because one sits under a loop and the other is under a function.


Overall, it is quite hard to achieve high efficiency when we
synthesize statements. The target codelet could appear anywhere in
the database program, and a sequence of statements tends to have more
references that need renaming. Therefore we expect synthesizing
statements to be a more difficult task. Although role matching cannot
be applied to synthesizing stand-alone functionality, it can still be
considered when the surrounding context is necessary for a codelet,
especially for an incomplete functionality.

\subsubsection{Code merging and Testing}

One problem with using the codelets from the database programs is that
the naming schemes are different from the ones in the original draft
program. Therefore, after we have completed the draft program, we
search for reference substitution such that the resulting program
refers back to the data defined in the draft program, which is quite
similar to code transplantation~\cite{Barr2015}.

\begin{wrapfigure}{R}{0.4\textwidth}
  \vspace{-20pt}
\begin{minipage}{0.4\textwidth}
\begin{algorithm}[H]
  \caption{merge}
  \begin{algorithmic}[1]
    \Require A completed draft program, $P_s \in \mathcal{L}$ and a database
    program $P_d \in \mathcal{L}$
    \Ensure A correct completion $P_c$

    \If {$\text{no\_undefined\_refs}(P_s)$}
      \If {is\_correct($P_s$)}
        \State \Return $P_s$
      \EndIf
      \State \Return $null$
    \EndIf

    \For {$u \gets \text{next\_undefined\_ref}(U)$}
      \For {$r \gets \text{next\_ref}(P_d)$}
        \If {same\_type($P_s, u, r$)}
          \State $P_s' \gets \text{substitute}(P_s, u, r)$
          \State $P_c \gets \text{merge}(P_s', P_d)$
          \If {$P_c \neq null$}
          \State \Return $P_c$
          \EndIf
        \EndIf
      \EndFor
    \EndFor
    \State \Return $null$
  \end{algorithmic}
  \label{alg:merge}
\end{algorithm}
\end{minipage}
\end{wrapfigure}

The algorithm is showed in Procedure~\ref{alg:merge}. The task here is
essentially searching for a mapping between the references across two
programs. We first check whether we have undefined references in the
program at line 1. If not, we check the program correctness against
the requirement at line 2. If it is correct, then we have a solution. If there
is still undefined reference in the program, we then try to rename
each undefined reference $u$ to another defined reference $r$ at line
10. We repeat by recursively calling itself until no more undefined
names exist in the program. We guide the search by using types. When
we are considering renaming $u$ to $r$, we rename $u$ only if their
types are the same. Again, types give us better efficiency. If at any point
the algorithm cannot rename a reference due to the lack of available
target references in another program, the algorithm will backtrack and
try another renaming for a previous reference.

This reference substitution step is performed every time we complete a
draft and thus the whole algorithm suffers from exponential
blowup. Since the expressions and statements are all from the database
programs which are finite, we have a finite search space. Moreover, we
also set a time limit on the entire search process and thus our main
synthesis algorithm is guaranteed to terminate.


After we have finished renaming all references in a completed program,
we validate the solution against of the requirement either in form of
a predefined input-output test suite or a predefined API call sequence
constraint given as a finite automaton. If users provide IO tests, we
run the solution on the provided test suite to validate its
correctness. If an API call sequence constraint is given instead, we
encode the constraint into Java source code in which API calls are
captured and new variables are defined to keep track of the current
state in the finite automaton. When the complete program is run, the
constraint will be automatically checked and thus the correctness is
determined. We also set a time limit for program execution to ensure
termination. Notice that we could let the synthesis algorithm produce
multiple solutions by letting it continue the search after a correct
completion is found. If there are multiple correct completions, we
will rank them in the order they appear and return as many solutions
as required.

Note that it is very easy to add a selection function to choose the
best solution among all the completions. This is quite useful if other
requirements not represented as tests or API call sequence constraints
are desirable. For example, we added a simple filter where we ignore
solutions with two consecutive and redundant return
statements. Potentially, one or more layers of selections could be
done after validation to ensure program properties.

\section{Evaluation}
\label{sec:eval}

\begin{figure*}
    \centering
    \begin{subfigure}[b]{0.45\textwidth}
      \begin{lstlisting}
/* COMMENT:
 * Longest common subsequence algorithm
 * TEST:
 * int[] s1 = {2, 6, 8, 0};
 * int[] s2 = {1, 6, 8, 10};
 * int[] expected = {6, 8};
 * __solution__
 * test_lcs(s1, s2, expected);
 */
void lcs(int[] X, int[] Y, int m, int n, int[] result) {
  // longest common subsequence
  int[][] L = new int[15][15];
  int i, j, index, s;
  // build the LCS table here
  ??
  // store the subsequence result
  index = L[m][n];
  s = index; i = m; j = n;
  while((i > 0) && (j > 0))
    if(X[(i - 1)] == Y[(j - 1)]) {
      result[(index - 1)] = X[(i - 1)];
      i --; j --; index --;
    } else {
      if(L[(i - 1)][j] > L[i][(j - 1)])
        i --;
      else
        j --;
    }
}
      \end{lstlisting}
      \vspace{-20pt}
      \caption{LCS with Customized Result Display}
    \end{subfigure}
    \begin{subfigure}[b]{0.45\textwidth}
      \begin{lstlisting}
/* COMMENT:
 * Setting up an HTTP server that serves
 * the content of a local file
 * TEST:
 * import com.sun.net.httpserver.*;
 * import java.io.OutputStream;
 * __solution__
 * int port = 23456;
 * HttpServer server = http("http_test.txt", port);
 * test_server(new URL("http://localhost:23456/"));
 * server.stop(0);
 */
public HttpServer http(String filename,
                       int port) {
  String content;
  // read the content of the file
  ??
  HttpServer server;
  HttpHandler handler = new HttpHandler() {
    public void handle(HttpExchange he) throws Exception {
      he.sendResponseHeaders(??,
        content.length());
      OutputStream os =
        he.getResponseBody();
      os.write(content.getBytes());
      os.close();
    }
  };
  // set up an http server
  ??
  return server;
}
      \end{lstlisting}
      \vspace{-20pt}
      \caption{Setting up an HTTP server that serves the content of a text file}
    \end{subfigure}
    \caption{Draft Programs Used for Experiments}
    \label{fig:sketches}
\end{figure*}

Our goal is to evaluate the performance of program splicing and its
ability to complete a draft program using a large code corpus such
that the resulting program meets a correctness requirement within a reasonable
amount of time. The experiment consists of completing a set of draft
programs given a code database where a set of relevant statistics for
each run is recorded. In addition, we test the performance of our code
search method afterwards and show the results of the user study we
conducted where we test whether our synthesis tool could increase
programming productivity. 

\subsection{Benchmarks}
%
In this section, we briefly describe our benchmark problems followed
by the experiments and the results. We evaluate the performance of
program splicing and select a set of benchmark problems with
corresponding draft programs to automate the process where users try
to bring external resources from the web and merge them into the
existing codebase. We ensure that the code we synthesize is quite
common in popular online code repositories so that it is more likely
to find them in our program database.

It is desirable to compare our method with existing synthesis methods
including Sketch~\cite{lezama06}, syntax-guided
synthesis~\cite{alur2015syntax}, code reuse tools such as
S${^6}$~\cite{reiss2009}, Code Conjure~\cite{hummel2008code},
CodeGenie~\cite{lazzarini2009applying} and Hunter~\cite{wang2016} or
other statistical methods~\cite{raghothaman2015swim,
  gvero2015interactive}. However, none of these methods are
comparable, because (1) traditional synthesis methods such as Sketch
do not search for or use existing source code, (2) code reuse methods only consider
programs at the granularity of functions
instead of more fine-grained level such as statements and expressions
and (3) some methods such as SWIM~\cite{raghothaman2015swim} and
anyCode~\cite{gvero2015interactive} only aim to synthesize
API-specific code snippets. Specifically, we fed the draft showed in
Figure~\ref{fig:binsearch-sketch} to Sketch and it was not able to
complete the draft within 30 minutes. In contrast, our splicing system
could generate the correct expressions within 5 seconds after the code
search is complete. Moreover, our splicing system could generate code
snippets while Sketch cannot handle statement synthesis problems.

\begin{wrapfigure}{r}{0.42\textwidth}
  \vspace{-10pt}
  \begin{lstlisting}
int binsearch(int[100] array, int x) {
  int low = 0, high = 99, result = -1;
	while(?? <= ??) {
    int mid = ?? / ??;
		if(array[mid] < x) {
		  low = mid + 1;
		} else if(array[mid] > x) {
		  high = ??;
		} else {
		  result = ??
			break;
		}
	}
	return result;
}
  \end{lstlisting}
\vspace{-20pt}
\caption{Binary Search Draft Fed to Sketch}
\label{fig:binsearch-sketch}
\vspace{-10pt}
\end{wrapfigure}

Code transplantation or $\mu$Scalpel~\cite{Barr2015} is actually the
most similar one to our work and we will use $\mu$Scalpel for
comparison with the correct donor programs provided to
$\mu$Scalpel. One thing to notice is that we cannot apply $\mu$Scalpel
to some of our system-related benchmark problems, because
$\mu$Scalpel targets at C programs instead of Java. Some system
programmings in C and in Java tend to be very different and therefore,
we only compare our tool with $\mu$Scalpel on some benchmark problems
where the differences of solutions are not significant.


Our benchmark problems consist of synthesizing components from online
repositories and we include 15 benchmark problems. The draft program
for most benchmark problem contains one or two statement hole and
expression holes. Each draft program has its own comments and
correctness requirements. Most benchmark problems use typical input-output tests except
for ``Echo Server'', ``Face Detection'' and ``Hello World GUI'' where
API call sequence constraint is used to check the correctness. Here,
we highlight two draft programs from the benchmark problems which are
showed in figure~\ref{fig:sketches}. All the draft programs are listed
in Appendix~\ref{subsec:draft}.

\begin{itemize}
\item \verb|LCS Table Building|: A user calculates the longest common
  subsequence of two integer arrays, and she has a written a draft
  program with the code snippets to extract the subsequence from the
  table and display the result. A hole is left for the code that
  builds the table for running dynamic programming algorithm.
\item \verb|HTTP Server|: A user would like to set up an HTTP server
  that serves the content of a text file. She wrote a draft program
  which has a HTTP request handler, but she does not remember how to
  read from a text file and how to set up an HTTP server. Two holes
  are left for the code that reads from a text file and the code that
  sets up an HTTP server. In addition, she also leaves a hole for the
  response status code in the request handler.
\end{itemize}

\newcommand{\vcell}[2][c]{%
  \begin{tabular}[#1]{@{}l@{}}#2\end{tabular}}
\begin{table*}
  \centering
    \begin{tabular}{| c | p{1.4cm} | p{1.4cm} | p{1.4cm} | p{1.0cm} | p{0.7cm} | p{0.7cm} | p{0.7cm} | p{1.2cm} |}
    \hline
    Benchmarks               & Synthesis Time    & No Roles          & No Types       & LOC   & Var & Holes (expr-stmt) & Test & $\mu$Scalpel \\ \hline

    Echo Server              & 3.0           & 4.0          & 17.1      & 9-17  & 1   & 1-1 & C & N/A \\ \hline
    Sieve Prime              & 4.6           & 33.0        & 8.8       & 12-17 & 2   & 2-1 & 3 & 162.1 \\ \hline
    Collision Detection      & 4.2           & 6.3          & 5.3        & 10-15 & 2   & 2-1 & 4 & N/A \\ \hline
    Collecting Files         & 3.0           & 6.0          & 27.0     & 13-25 & 2   & 1-1 & 2 & timeout \\ \hline
    Face Detection           & 8.1          & 12.2        & 43.1     & 21-28 & 2   & 1-1 & C & N/A \\ \hline
    Binary Search            & 15.4   & 16.0        & 47.9     & 12-20 & 5   & 1-1 & 3 & timeout \\ \hline
    Hello World GUI          & 16.0         & timeout          & timeout        & 24-33 & 4   & 1-2 & C & N/A \\ \hline
    HTTP Server              & 41.1         & 87.4          & timeout        & 24-45 & 6   & 1-2 & 2 & N/A \\ \hline
    Prim's Distance Update   & 61.1        & 66.4       & timeout        & 53-58 & 11  & 1-1 & 4 & timeout \\ \hline
    Quick Sort               & 77.2         & 191.5      & 217.6    & 11-18 & 6   & 1-1 & 1 & timeout \\ \hline
    CSV                      & 88.4        & timeout          & timeout        & 13-23 & 4   & 1-2 & 2 & timeout \\ \hline
    Matrix Multiplication    & 108.9        & 151.9       & timeout        & 13-15 & 8   & 1-1 & 1 & timeout \\ \hline 
    Floyd Warshall           & 110.4   & timeout          & timeout        & 9-12  & 7   & 1-1 & 7 & timeout \\ \hline
    HTML Parsing             & 140.4       & timeout          & timeout        & 20-34 & 5   & 1-2 & 2 & N/A \\ \hline
    LCS                      & 161.5       & 168.8      & timeout        & 29-36 & 10  & 0-1 & 1 & timeout \\ \hline

    \hline
    \end{tabular}
  \caption{Benchmarks. ``C'' in the ``Test'' column indicates an API call sequence constraint is used to check the correctness}
  \label{tbl:benchmarks}
  \vspace{-20pt}
\end{table*}

\subsection{Experiments}

We implemented program splicing in Scala 2.12.1 based on 64-bit
OpenJDK 8 and we used BeanShell~\cite{beanshell} and
Nailgun~\cite{nailgun} to test all the completed draft programs. For
each benchmark problem, we ran the system on the draft program we
derived. These experiments were conducted on a 2.2GHz Intel Xeon CPU
with 12 cores and 64GB RAM. For each program, we set the time limit to
5 minutes and record the runtime for synthesis. To roughly have a
sense of the search space size, we list the number of variables and
holes in each draft program, the line number and the number of
database programs we use for synthesis. Finally, we list the LOC of
the draft program and its completed version. Our corpus comes from
\verb|Maven 2012| dataset from Sourcerer~\cite{sajnani:icsem2014,
  Ossher:WCRE12, Bajracharya:SCP14}. We extracted over 3.5 million
methods with features from this corpus.


\subsubsection{Synthesis Algorithm Evaluation}

Table~\ref{tbl:benchmarks} shows the results for each benchmark problem
with $k = 5$ where $k$ is the number of database programs we
retrieve. We set $k = 5$ because five programs are usually sufficient
to ensure that the retrieved programs contain the target codelet we
want to synthesize. In addition, we put more weights on
features that consider comments and variable names to search the
database $k$-nearest-neighbor search. The choice on weight selection
is explained in section~\ref{subsection:search}.

According the results showed in Table~\ref{tbl:benchmarks},
data-driven synthesis works for all benchmark problems. The time
required for most code search which is based on k-nearest-neighbor
search is approximately 15 seconds meaning that the code search is
very efficient, given that we have millions of functions in the
database. For most of the benchmark problems, our method was able to
complete the draft program under two minutes and the number of tests
required is no more than five, indicating that users of our system do not
have the burden of writing too many tests. Notice that for ``Echo
Server'', ``Face Detection'' and ``Hello World GUI'', a letter ``C''
is used to signal an API call sequence constraint being used to test the
correctness. Because of the Java testing infrastructure we used for
testing, a large amount of program execution overhead was reduced. We
can also see that synthesis takes more time as the number of holes and
the number of variables increase. Having more holes, more variables
and sometimes more lines leads to larger combinatorial search space
for hole substitutions with codelets, and more variables increase the
search space for code merging and renaming.



\textbf{Impact of type matching and role matching} Types guide the
search during hole substitutions and code merging and it potentially
eliminates final solutions that do not type check. In addition, role
matching eliminates the expression substitutions where the role of a
candidate expression is different from the role of a hole. In order to
understand their impacts, we record the synthesis time without using
types as heuristic which is showed in the ``No Types'' column of
Table~\ref{tbl:benchmarks}. ``No Roles'' column shows the runtime where
role matching heuristic is not applied. We can see that using types and roles
can reduce a large amount of search space, although types seem to be
more effective. These heuristics become more and more important for
larger draft programs as the number of variables increases. Without
types and role matching, our synthesis algorithm even timed out for
some harder benchmark problems. Notice that role matching is applied
when we synthesize for expressions, as we cannot apply role matching
when synthesizing statement sequences and thus we do not see any
difference in the ``LCS'' benchmark problem.

\textbf{$\mu$Scalpel Comparison} Code transplantation~\cite{Barr2015}
is very similar to our work except for the fact that they do not
consider using a large code corpus. However, it is still worthwhile to
conduct a series of performance comparisons since $\mu$Scalpel also
extracts code snippets from external programs or \emph{donor}
programs. We ran $\mu$Scalpel on some of our benchmark
problems with correct donors specified for multiple times. Notice that
$\mu$Scalpel has some advantage over our system under this setting,
because $\mu$Scalpel does not need to search for a set of relevant
programs from a large code corpus. Nevertheless, even with such
advantage, most of the runs could not finish within 5 minutes except
for ``Sieve Prime'' which is one of the easy ones. Even though we did
not run $\mu$Scalpel on all benchmark problems, it is reasonable to
believe that the performance of $\mu$Scalpel which is based on genetic
programming is not as efficient as our system which is based on
enumerative search.


\subsubsection{Code Search Evaluation}
The synthesis result actually depends heavily on the quality of the
programs retrieved from the database. A \emph{high-quality} program
should contain the exact codelet we want to synthesize. Without
high-quality program, no solutions will exist in the search
space. Therefore, it is important to study the quality of the database
programs we used during synthesis and calculate the proportion of
high-quality programs. We define this quantity as \emph{precision}
which is equal to the number of high-quality programs versus the total
number of programs we used for synthesis.

\begin{wrapfigure}{r}{0.7\textwidth}
\vspace{-10pt}
\includegraphics[scale=0.3]{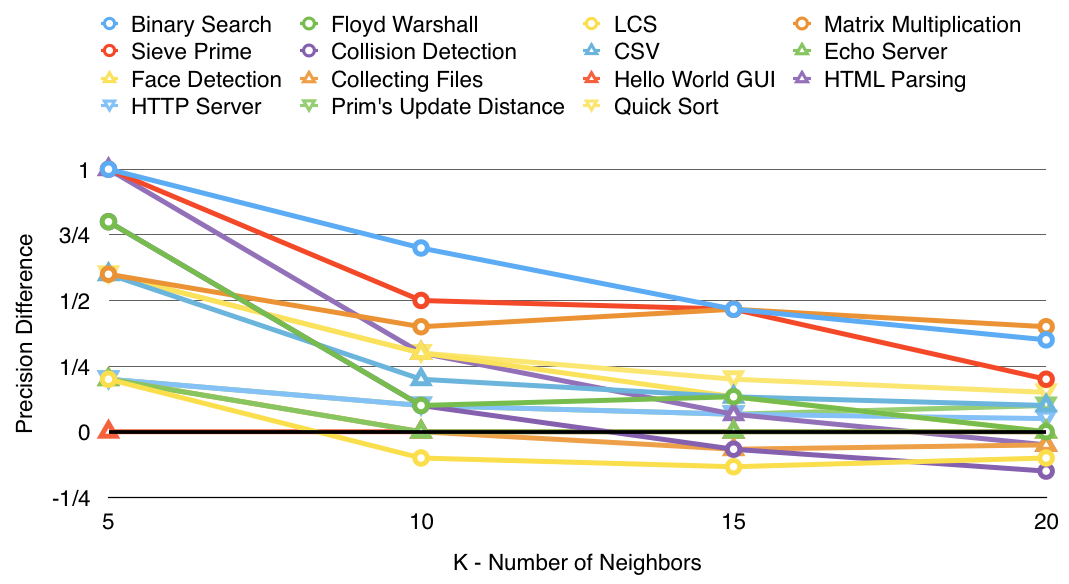}
\vspace{-20pt}
\caption{Our Code Search Precision Minus GitHub's Precision}
\label{fig:precision}
\vspace{-10pt}
\end{wrapfigure}

Technical difficulty exists when we try to calculating the precision,
because it requires deep analysis to check whether a portion of the
database programs satisfies a certain property which is quite
expensive. Searching for a codelet which satisfies a specification in
a program is essentially another expensive code search and
verification problem. Therefore, we come up with a good proxy to
approximate precision. In order to see whether a program has the
codelet we want, we can run the synthesis algorithm using that
particular program and remove time constraint. If a particular
database program has the codelet we want, eventually the synthesis
would be successful, because the search space is finite. Essentially,
our synthesis algorithm searches each database program for the target
codelet and checks its property using correctness requirements.

We calculate the precision as the number of database programs, $k$,
increases. We record the number of high-quality programs versus the
total number of programs we used for synthesis. Again, we put more
weights on natural language features and less on other features for
$k$-nearest-neighbor search. In addition, it is interesting to see
whether well-known code search engines could be beneficial to our
synthesis algorithm as well. Therefore, we compare our code search
method with GitHub~\cite{github} code search engine. We choose
GitHub because its search method is quite similar and comparable in
our case. Other code search methods which rely on code pattern or
other syntax element~\cite{jiang2007deckard, keivanloo2014spotting}
are quite different from our method and thus they are not comparable.
We search GitHub for similar programs using comments and keywords from
the draft programs and test their precision. We searched for programs
contained in files with \verb|.java| extension and also tagged as written
in Java language. We used ``best match'' ordering and selected top-$k$
entries from the search result. When there are multiple functions in a
source code, we simply pick the function that is most likely to
contain the target codelets.

Figure~\ref{fig:precision} shows the numerical difference between the
precision of our code search method and the precision from GitHub
search engine with various $k$ as the x-axis. Positive percentage
indicates our code search method finds more high-quality programs and
negative percentage means the opposite. From the figure, we can see
that our code search method is actually quite effective at fetching
high-quality programs across different benchmark problems with various
$k$. Moreover, our code search method can always find more
high-quality programs than well-known code search engine within five
programs, meaning that our method has better precision. In some cases
such as ``Collision Detection'' and ``LCS'', GitHub tends to be more
precise. After inspecting the programs from our database and the ones
from Github, we believe the reason is that our database happens to
have more repetitive and testing functions in the ``LCS'' case and
different implementations in the ``Collision Detection'' case.

As we were trying to calculate the precision, we discovered that
finding high-quality programs is actually quite difficult, because
even a small change could completely change the usability of a
program. In addition to the programs which contains syntactic features
that our system does not support, the following factors can influence
programs' usability, even for a programmer:

\textbf{Different implementation} If an algorithm has more than one
implementation, it is possible that we might retrieve another
implementation which cannot be used for synthesis. In most cases,
different implementations tend to not include the codelets we want to
synthesize. Suppose we want to synthesize the main loop of an
iterative binary search, it is impossible to use the recursive
version. Moreover, multi-threaded versions or other versions
that use different libraries are also quite popular and we cannot use
those as well.

In addition, some database programs might use different constants for
variable initializations and array sizes. Loops range might be
slightly different as well. Usually these differences lead to crashes
or logic error if we use those codelets to complete the draft without
any modification. Typically in this scenario it is trivial for the
users to change those constants and operators so that the retrieved
programs could be used. Therefore, we performed simple transformations
on the database programs to reflect that in order to prevent precision
from being undermined.



\textbf{Repetitive, irrelevant and invalid programs} With an enormous
amount of programs in a database, ensuring the quality of every single
program is difficult. It is quite easy to get garbage programs when
searching a large corpus, let alone the fact that we are using natural
language which is ambiguous by nature. For example, when we search for
quick sorts, we get back some other sort functions, driver functions
for sorting algorithms and also test functions with ``sort'' as part
of the name. Furthermore, repetitive and empty functions are also
quite popular. Repetitive programs might come from repository cloning
and duplicate code commit. Empty functions seem to be created and
abandoned later by programmers.

\begin{figure*}
    \centering
    \begin{subfigure}[b]{0.45\textwidth}
      \begin{lstlisting}
/**
 * TODO 1:
 * Use Sieve of Eratosthenes to test
 * primality of the given integer.
 */
static boolean sieve(int n) {
  boolean[] primes = new boolean[100];
  return primes[n];
}
/**
 * TODO 2:
 * Test the sieve of Eratosthenes
 * you've just written. Make sure to
 * test the program with the following
 * inputs:
 * n = {1, 2, 3, ..., 73}
 * Return true if the program is correct.
 * Otherwise, return false.
 */
static public boolean test() {
  return false;
}
      \end{lstlisting}
      \vspace{-20pt}
      \caption{Skeleton Program with Tests}
      \label{fig:user-study-skeleton}
    \end{subfigure}
    \begin{subfigure}[b]{0.45\textwidth}
      \begin{lstlisting}
/**
 * TODO 1:
 * Complete the following function.
 * You could use our system by replacing
 * the contents in COMMENT and TEST.
 * COMMENT:
 * Replace this comment with something
 * related to sieve of Eratosthenes
 * algorithm. 
 * TEST:
 * // test the sieve of Eratosthenes
 * // program by writing Java program here
 * __solution__
 * return(true);
 */
static boolean sieve(int n) {
  boolean[] primes = new boolean[100];
  ??
  return primes[n];
}
      \end{lstlisting}
      \vspace{-20pt}
      \caption{Draft Program}
      \label{fig:user-study-draft}
    \end{subfigure}
    \caption{\texttt{Sieve of Eratosthenes} tasks for the user study.}
    \vspace{-10pt}
\end{figure*}

\subsubsection{User Study}
It is unclear whether our system will be beneficial in the wild, for use by actual developers.  In this subsection,
we describe a user study aimed at answering this question.

\vspace{5 pt}
\noindent
\textbf{Study setup.}
We recruited 12 graduate students and
six professional programmers and developed four programming problems.  Each 
participant was asked to complete all four programming problems
using a web-based programming environment.   Per person, two problems were completed
using program splicing (we subsequently call this a ``with'' task),
and two without (a ``without'' task).  ``With'' and ``without'' tasks were assigned
to participants randomly.

In order to simulate the industrial programming setting where an engineer is
asked to develop a code meeting a provided specification, for each task, participants
were given a description of the target program they need to implement,
and also a description of the test cases they need to
write to verify the correctness of the program.
Figure~\ref{fig:user-study-skeleton} shows an example skeleton
program for the ``without'' task on the \text{Sieve of Eratosthenes} programming problem,
and
figure~\ref{fig:user-study-draft} shows a draft program where participants
need to put in comments and requirements to complete the ``with'' task. 

When completing both ``with'' and ``without'' tasks, participants
were encouraged to find and use relevant code snippets
from the Internet.  For the ``with'' tasks,  participants were asked to use our system to 
provide at least one candidate solution to the programming problem, but then they 
could choose to use that candidate, or not use it.
Before using the web-based programming environment and our
system, they were asked to finish a warm-up problem in order to be
familiar to the programming environment and our system.

To evaluate whether our system could boost programming productivity,
we recorded the amount of time the participants used to correctly complete each
programming problem. 
In order to determine whether there is a statistically
significant difference in task completion time for ``with'' versus ``without'' tasks for the same programming problem,
we define the following null hypothesis:

\vspace{5 pt}
\begin{center}
$H_0^P$ = ``For programming problem $P$, the expected `without' task completion time is
no greater than the expected `with' task completion time.''
\end{center}

\vspace{5 pt}
If this hypothesis is rejected at a sufficiently small $p$-value for a specific programming problem, it means that it is likely
that the average completion time is smaller for the ``with'' task than the ``without'' task, and hence program splicing likely
has some benefit on the problem.

Given the times recorded over each problem and each task, we use bootstrap\cite{efron1982jackknife} to calculate the $p$-value
for each programming problem. The bootstrap works by simulating a large number
of datasets from the original data by re-sampling with replacement
many times, and the $p$-value is approximated by the fraction of the time
when the null hypothesis holds in the simulated data sets.

In addition to measuring time, we also recorded the number of times that
``with'' task participants for each problem asked the program splicing system
for help.
Typically the
participants would stop using our system after they have received a
useful codelet, and so a large number of requests may indicate an inability
of the system to produce a useful result.

\vspace{5 pt}
\noindent
\textbf{Programming Problems.}
The four programming problems were as follows:
\begin{itemize}
\item \texttt{Sieve of Eratosthenes}: Implement the Sieve of
  Eratosthenes to test the primality of an integer. This is an
  interesting programming problem because it is purely algorithmic,
  involves no systems programming or API calls, and further codes to
  solve this problem are ubiquitous on the Internet.  Going into our
  study, we expected program splicing to be of little use on this
  problem, because an Internet search should result in many different
  Sieve programs which should be trivial to tailor into a solution to
  the problem.  Given this and the fact that test codes are so easy to
  write, we expected participants will use the least amount of time to
  finish this problem, regardless of whether they are given a ``with''
  or ``without'' task for this problem.
\item \texttt{File Name Collection}: Collect all file names under a
  directory tree recursively and return a list of file names. We chose
  this problem because it represents an easy systems programming
  problem. Further, there is no standard solution to this problem,
  while it is still quite easy to write tests.  Therefore we expected
  Internet search to be less useful, whereas program splicing might be
  quite helpful.
\item \texttt{CSV Matrix Multiplication}: Read a matrix from a CSV
  file, square the matrix and return it as a 2d-array. This problem
  includes a combination of system programming and algorithmic
  programming. We chose this problem expecting that ``with'' task
  programmers would need to use our system multiple times in an
  interactive manner to generate two independent code snippets. Given
  this, we expected that the time gap between the ``with'' task and
  ``without'' task participants to be smaller.
\item \texttt{HTML Parsing}: Read and parse an HTML document from a
  text file, store all links which contain a given word into a result
  list and return the result list. This is the most difficult problem
  among the four. Not only would those ``with'' task participants need
  to use our system multiple times, but they are required to write
  test for HTML manipulation. Specifically, program splicing
  necessitates that participants manually provide HTML to build test
  cases that are used to validate the correctness of the code for
  extracting links from the parsed HTML document. At the same time,
  the JSoup~\cite{jsoup} HTML parsing library that we asked
  participants to use has rather comprehensive and straight-forward
  documentation. Hence, we expected that time gap between ``with'' and
  ``without'' task participants would be the smallest among the four
  problems.
\end{itemize}

\begin{wrapfigure}{r}{0.5\textwidth}
  \vspace{-10pt}
  \includegraphics[scale=0.55]{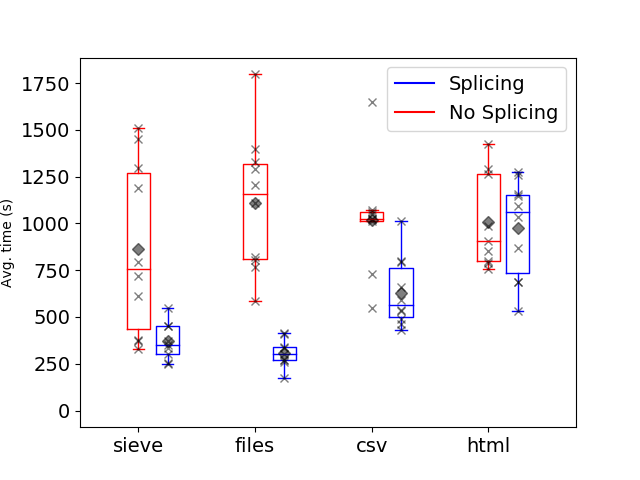}
  \vspace{-10pt}
  \caption{Time spent on each programming problem, with and without splicing.}
  \label{fig:user-mean}
  \vspace{-10pt}
\end{wrapfigure}

\vspace{5 pt}
\noindent
\textbf{Results.}
Figure~\ref{tbl:user-study-pval} shows the $p$-values for each
programming problem, as well as the number of times code splicing was
invoked for each problem's ``with'' task. Figure~\ref{fig:user-mean}
shows time spent on each submission with and without splicing,
including the average time and the box plots. We can see that for most
programming problems except for \verb|HTML|, the average time used to
finish the ``with'' task is is significantly lower than the time
required to finish the ``without'' task.  The $p$-values in
figure~\ref{tbl:user-study-pval} are also small enough for us to
reject the null hypotheses (stating that there is no utility to
program splicing) with over 99\% confidence.

Note that the average number of program splicing invocations for most
problems (except \texttt{HTML Parsing}) is very close to one, meaning
that program splicing could return codelets that the participants
could use to complete the problem with only one try. We argue that
this also indicates that the system is rather easy to use, and is
indeed able to boost programming productivity in many cases. As the
level of difficulty of the problem increases, so does the benefit of
using our system.

It is, however, useful to consider the \texttt{HTML Parsing}
programming problem, which is the one case where program splicing was
not useful.  Why is this?  What is the failure mode? After careful
investigation, we believe that there are two reasons program splicing
did not help. First, the documentation of the HTML parsing library
used, \verb|JSoup|~\cite{jsoup}, is very comprehensive and
well-done. Hence the problem was easy. Second, it is very easy to make
mistakes when writing tests, which require developing correct HTML
code and inserting it in a test. We found that participants had a
difficult time with this. Participants typically forgot to escape
quote characters within a string when loading a variable containing
even very simple HTML. Although providing a better programming
environment would be very helpful, the difficulty in writing tests
meant that program splicing was less helpful. That said, writing tests
has independent value, and if the difficulty in writing tests was the
key impediment to using splicing, it may not be a strong argument
against the tool.

\begin{wrapfigure}{r}{0.45\textwidth}
  \begin{tabular}{| c | c | c |}
    \hline
    Problem  & $p$-value & Avg. Number of Invocations\\ \hline
    Sieve & 0.00008   & 1.2 \\ \hline
    CSV   & 0.0002       & 1.2 \\ \hline
    Files & 0       & 1 \\ \hline
    HTML  & 0.5   & 2.45 \\ \hline
    \hline
  \end{tabular}
  \caption{$p$-value at which the null hypothesis is rejected, and the
    average number of program splicing invocations for each
    programming problem.}
  \label{tbl:user-study-pval}
  \vspace{-10pt}
\end{wrapfigure}

We close this subsection by asking: When is program splicing likely
most useful for programmers? One surprising case seems to be
programming problems that are deceptively simple, containing intricate
algorithmics (loops and recursion) that programmers tend to have a
difficult time with. \texttt{Sieve of Eratosthenes} falls in this
category.  The Sieve appears to be very simple, and so we initially
expected splicing to be of little use.   However, due to the perceived
simplicity, we found that ``without'' participants tended to write
their own solutions without consulting the Internet (even though we
encouraged Internet use)---and this over-confidence resulted in buggy
programs and longer development times. We were especially surprised to
find this in the case of \texttt{Sieve of Eratosthenes} and also in
the case of matrix multiplication part in \texttt{CSV}, where there
are many solutions available on the Internet. Use of program splicing
protected ``with'' participants from such difficulties.

We also found splicing to be useful when documentation is lacking and
there is not a standard way of doing things. Consider \texttt{CSV} and
\texttt{Collecting File Names} where the official Java documentation does
not provide any code snippets on how to parse a CSV files or how to
collect file names under a directory subtree. ``Without'' participants
had to rely on combing through solutions from
StackOverflow~\cite{stackoverflow}, where multiple solutions exist,
using different libraries, each with various pros and cons. Program
splicing cuts out the need for manual searching and understanding many
different possible solutions---if the splicing succeeds and passes the
provided test cases, the user can be relatively confident that the
provided solution is correct.


\newcommand{\gtnote}[1]{\textbf{[FIXME: [#1]]}}

\section{Related Work}
\label{sec:relwork}

The problem of program synthesis has been considered for a long
time~\cite{Pnueli1989,Alur2015} and recently it has been successfully
applied to domain-specific tasks~\cite{yaghmazadeh2016, feng2017,
  feng2016, polozov2015flashmeta}. In particular, the notion of drafts
used in program splicing is inspired by ``sketches''~\cite{lezama06}
and ``templates'' from previous approaches to
synthesis~\cite{Srivastava2012}. The sort of combinatorial search that
our synthesis algorithm uses has parallels in enumerative approaches
to synthesis~\cite{udupa2013transit,Feser2015}. However, the key
difference between program splicing and other synthesis techniques is
that our method reuses existing source code from the web instead of
generating programs from scratch, and this gives our method better
scalability advantage and traditional synthesis methods such as
Sketch~\cite{lezama06} lack the capability of synthesizing
\emph{statements} efficiently.

Code transplantation and other methods that use genetic
programming~\cite{Barr2015, petke2014using, harman2014babel,
  jia2015grow, marginean2015automated} are very similar to our
work. They transplant external code snippets, functionalities or
organs across multiple programs. However, they do not considering
searching a large code corpus, and genetic-programming based methods
suffer from a serious
efficiency issue according to our experiment
results. CodePhage~\cite{sidiroglou2015automatic} also transplants
arbitrary code snippets across different applications, but the
transplantaions are done exclusively for binary programs. In addition,
their experiments only consider adding checks for program repair, and
it is still unclear whether it is applicable to arbitrary code.

Code reuse tools such as CodeConjure~\cite{hummel2008code},
S${^6}$~\cite{reiss2009}, CodeGenie~\cite{lazzarini2009applying}, and
Hunter~\cite{wang2016} are quite beneficial to programmers. It has
been showed that these tool could increase the efficiency of
programming. Some are similar to our work in which they use natural
language and tests to search for relevant programs. However, the major
difference between these tools and ours is that they usually consider
programs at the granularity of functions and sometimes a piece of type adapter
code is required~\cite{wang2016}. On the other hand, our method
provides a Sketch-like interface where programmers can write
holes. Naturally, our method needs to dig into functions and look for
statements and expressions, which opens a door to an exponentially
larger code database for reuse.

Program analysis and synthesis using a large pool of existing source
code, or ``Big Code'', recently has gained a lot of attention. Mishne
et al.~\cite{mishne12} focuses on mining the specification for API
calls from a large amount of code snippets. Statistical methods such
as graphical models~\cite{Raychev2015}, language
models~\cite{Raychev2014, hindle2012naturalness, nguyen2015graph} and
learning from noisy data~\cite{raychev2016} have been showed to be
quite effective in inferring program properties and code
completion. Our work, however, does not depend heavily on statistical
methods. SWIM~\cite{raghothaman2015swim} and other
works~\cite{gvero2015interactive} also uses natural language query and
the web to search for code snippets such as API usages, but they do
not consider draft programs that offer a context with which new
programs must be merged. DeepCoder~\cite{balog2016deepcoder} has
gained a lot of popularity recently. They successfully applied deep
learning to the problem of program synthesis and show that deep
learning could be beneficial in reducing the search space during
synthesis. Our work is different in the sense that we do not use any
statistical method, and DeepCoder works on a simpler language.

Copy-and-paste has long been seen to be a problematic approach to
programming. However, a recent paper~\cite{narasimhan2015copy} seeks
to redeem copy-and-paste through a method that finds clones of a fully
written program and automatically merges these cloned code with the
new code. However, their work only considers extremely similar
programs whose parse trees are a few edits away and does not consider
draft programs. In contrast, our approach can bring arbitrary programs
from the internet into a programming context, and uses a combinatorial
synthesizer to splice this code into the
context. Gilligan~\cite{Holmes2013} serves as a code reuse assistant,
which is very similar to our approach. However, it does not consider
using external source code.

Much of the related work on finding similar code has focused on {\em
  clone detection}: finding syntactically exact or nearly-exact copies
of source code fragments. See \cite{roy2009comparison} for a
relatively recent survey of these techniques. Code clone detection
usually assumes that two fragments of code were derived from the same
original code, flagging fragments with certain types or quantities of
edits between them as clones. Syntax elements have been shown to be
effective in code search~\cite{jiang2007deckard, keivanloo2014spotting},
but our code search mainly relies on natural language with minor
syntax elements. We also differ from other token-based code search engines
in that we do not only rely on natural language tokens in code.


Code search has been performed in various other settings, using
different code features: TRACY~\cite{david2014tracelet} uses k-limited
static program paths (called tracelets) to search similar stripped
binary code snippets, Exemplar~\cite{Grechanik2010} uses Java APIs
exercised to aid code search, Portfolio~\cite{McMillan2011} uses
function call graph to improve code search and ranking. Refer Table 3
in~\cite{McMillan2011} for a comprehensive comparison of various code
search engines: most of these are targeted towards creating
user-facing code search engines that can find relevant code based on a
user-specified query. SMT solves~\cite{stolee2012toward} have also been used
for semantic code search and the combination of SMT solvers and
semantic code search has been shown to be effective at repairing
programs~\cite{ke2015repairing}.

\section{Conclusion}
\label{sec:conclusion}
In this paper, we introduce {\em program splicing}, a synthesis-based
approach to programming that can serve as a principled and automated
substitute for copying and pasting code from the internet. The main
technology behind program splicing is a program synthesizer that can
query a database containing a large number of code snippets mined from
open-source software repositories. Our experiments show that it is
possible to synthesize code snippets by fruitfully combining such
database queries with combinatorial exploration of a space of
expressions and statements. We also conducted a user study and the
results show that our method could indeed boost programming productivity.


One important future work is to ensure the high quality of database
programs, because the effect on the quality of the database program
could be very significant. We could develop better features and
similarity metrics in order to increase the code search precision.

\begin{acks}                            
  This material is based upon work supported by the
  \grantsponsor{GS100000001}{National Science
    Foundation}{http://dx.doi.org/10.13039/100000001} under Grant
  No.~\grantnum{GS100000001}{nnnnnnn} and Grant
  No.~\grantnum{GS100000001}{mmmmmmm}.  Any opinions, findings, and
  conclusions or recommendations expressed in this material are those
  of the author and do not necessarily reflect the views of the
  National Science Foundation.
\end{acks}

\bibliography{refs}


\begin{thebibliography}{00}


\ifx \showCODEN    \undefined \def \showCODEN     #1{\unskip}     \fi
\ifx \showDOI      \undefined \def \showDOI       #1{{\tt DOI:}\penalty0{#1}\ }
  \fi
\ifx \showISBNx    \undefined \def \showISBNx     #1{\unskip}     \fi
\ifx \showISBNxiii \undefined \def \showISBNxiii  #1{\unskip}     \fi
\ifx \showISSN     \undefined \def \showISSN      #1{\unskip}     \fi
\ifx \showLCCN     \undefined \def \showLCCN      #1{\unskip}     \fi
\ifx \shownote     \undefined \def \shownote      #1{#1}          \fi
\ifx \showarticletitle \undefined \def \showarticletitle #1{#1}   \fi
\ifx \showURL      \undefined \def \showURL       #1{#1}          \fi
\providecommand\bibfield[2]{#2}
\providecommand\bibinfo[2]{#2}
\providecommand\natexlab[1]{#1}
\providecommand\showeprint[2][]{arXiv:#2}

\bibitem[\protect\citeauthoryear{??}{git}{2016}]%
        {github}
 \bibinfo{year}{2016}\natexlab{}.
\newblock \bibinfo{title}{GitHub}.
\newblock   (\bibinfo{year}{2016}).
\newblock
\showURL{%
\url{https://github.com}}
\newblock
\shownote{Accessed: 2016-08-25.}


\bibitem[\protect\citeauthoryear{??}{bea}{2017}]%
        {beanshell}
 \bibinfo{year}{2017}\natexlab{}.
\newblock \bibinfo{title}{BeanShell}.
\newblock   (\bibinfo{year}{2017}).
\newblock
\showURL{%
\url{https://github.com/beanshell/beanshell}}
\newblock
\shownote{Accessed: 2017-04-04.}


\bibitem[\protect\citeauthoryear{??}{jso}{2017}]%
        {jsoup}
 \bibinfo{year}{2017}\natexlab{}.
\newblock \bibinfo{title}{JSoup}.
\newblock   (\bibinfo{year}{2017}).
\newblock
\showURL{%
\url{https://jsoup.org}}
\newblock
\shownote{Accessed: 2017-04-02.}


\bibitem[\protect\citeauthoryear{??}{nai}{2017}]%
        {nailgun}
 \bibinfo{year}{2017}\natexlab{}.
\newblock \bibinfo{title}{Nailgun}.
\newblock   (\bibinfo{year}{2017}).
\newblock
\showURL{%
\url{http://martiansoftware.com/nailgun/}}
\newblock
\shownote{Accessed: 2017-04-04.}


\bibitem[\protect\citeauthoryear{??}{sta}{2017}]%
        {stackoverflow}
 \bibinfo{year}{2017}\natexlab{}.
\newblock \bibinfo{title}{Stackoverflow}.
\newblock   (\bibinfo{year}{2017}).
\newblock
\showURL{%
\url{https://stackoverflow.com}}
\newblock
\shownote{Accessed: 2017-04-02.}


\bibitem[\protect\citeauthoryear{Alur, Bodik, Juniwal, Martin, Raghothaman,
  Seshia, Singh, Solar-Lezama, Torlak, and Udupa}{Alur et~al\mbox{.}}{2015a}]%
        {Alur2015}
\bibfield{author}{\bibinfo{person}{Rajeev Alur}, \bibinfo{person}{Rastislav
  Bodik}, \bibinfo{person}{Garvit Juniwal}, \bibinfo{person}{Milo~MK Martin},
  \bibinfo{person}{Mukund Raghothaman}, \bibinfo{person}{Sanjit~A Seshia},
  \bibinfo{person}{Rishabh Singh}, \bibinfo{person}{Armando Solar-Lezama},
  \bibinfo{person}{Emina Torlak}, {and} \bibinfo{person}{Abhishek Udupa}.}
  \bibinfo{year}{2015}\natexlab{a}.
\newblock \showarticletitle{Syntax-guided synthesis}.
\newblock \bibinfo{journal}{{\em Dependable Software Systems Engineering\/}}
  \bibinfo{volume}{40} (\bibinfo{year}{2015}), \bibinfo{pages}{1--25}.
\newblock


\bibitem[\protect\citeauthoryear{Alur, Bodik, Juniwal, Martin, Raghothaman,
  Seshia, Singh, Solar-Lezama, Torlak, and Udupa}{Alur et~al\mbox{.}}{2015b}]%
        {alur2015syntax}
\bibfield{author}{\bibinfo{person}{Rajeev Alur}, \bibinfo{person}{Rastislav
  Bodik}, \bibinfo{person}{Garvit Juniwal}, \bibinfo{person}{Milo~MK Martin},
  \bibinfo{person}{Mukund Raghothaman}, \bibinfo{person}{Sanjit~A Seshia},
  \bibinfo{person}{Rishabh Singh}, \bibinfo{person}{Armando Solar-Lezama},
  \bibinfo{person}{Emina Torlak}, {and} \bibinfo{person}{Abhishek Udupa}.}
  \bibinfo{year}{2015}\natexlab{b}.
\newblock \showarticletitle{Syntax-guided synthesis}.
\newblock \bibinfo{journal}{{\em Dependable Software Systems Engineering\/}}
  \bibinfo{volume}{40} (\bibinfo{year}{2015}), \bibinfo{pages}{1--25}.
\newblock


\bibitem[\protect\citeauthoryear{Bajracharya, Ossher, and Lopes}{Bajracharya
  et~al\mbox{.}}{2014}]%
        {Bajracharya:SCP14}
\bibfield{author}{\bibinfo{person}{Sushil Bajracharya}, \bibinfo{person}{Joel
  Ossher}, {and} \bibinfo{person}{Cristina Lopes}.}
  \bibinfo{year}{2014}\natexlab{}.
\newblock \showarticletitle{Sourcerer: An infrastructure for large-scale
  collection and analysis of open-source code}.
\newblock \bibinfo{journal}{{\em Science of Computer Programming\/}}
  \bibinfo{volume}{79} (\bibinfo{year}{2014}), \bibinfo{pages}{241 -- 259}.
\newblock
\showISSN{0167-6423}
\showDOI{%
\url{http://dx.doi.org/10.1016/j.scico.2012.04.008}}


\bibitem[\protect\citeauthoryear{Balog, Gaunt, Brockschmidt, Nowozin, and
  Tarlow}{Balog et~al\mbox{.}}{2016}]%
        {balog2016deepcoder}
\bibfield{author}{\bibinfo{person}{Matej Balog}, \bibinfo{person}{Alexander~L
  Gaunt}, \bibinfo{person}{Marc Brockschmidt}, \bibinfo{person}{Sebastian
  Nowozin}, {and} \bibinfo{person}{Daniel Tarlow}.}
  \bibinfo{year}{2016}\natexlab{}.
\newblock \showarticletitle{DeepCoder: Learning to Write Programs}.
\newblock \bibinfo{journal}{{\em arXiv preprint arXiv:1611.01989\/}}
  (\bibinfo{year}{2016}).
\newblock


\bibitem[\protect\citeauthoryear{Barr, Harman, Jia, Marginean, and Petke}{Barr
  et~al\mbox{.}}{2015}]%
        {Barr2015}
\bibfield{author}{\bibinfo{person}{Earl~T. Barr}, \bibinfo{person}{Mark
  Harman}, \bibinfo{person}{Yue Jia}, \bibinfo{person}{Alexandru Marginean},
  {and} \bibinfo{person}{Justyna Petke}.} \bibinfo{year}{2015}\natexlab{}.
\newblock \showarticletitle{Automated Software Transplantation}. In
  \bibinfo{booktitle}{{\em Proceedings of the 2015 International Symposium on
  Software Testing and Analysis}} {\em (\bibinfo{series}{ISSTA 2015})}.
  \bibinfo{publisher}{ACM}, \bibinfo{address}{New York, NY, USA},
  \bibinfo{pages}{257--269}.
\newblock
\showISBNx{978-1-4503-3620-8}
\showDOI{%
\url{http://dx.doi.org/10.1145/2771783.2771796}}


\bibitem[\protect\citeauthoryear{David and Yahav}{David and Yahav}{2014}]%
        {david2014tracelet}
\bibfield{author}{\bibinfo{person}{Yaniv David} {and} \bibinfo{person}{Eran
  Yahav}.} \bibinfo{year}{2014}\natexlab{}.
\newblock \showarticletitle{Tracelet-based code search in executables}. In
  \bibinfo{booktitle}{{\em ACM SIGPLAN Notices}}, Vol.~\bibinfo{volume}{49}.
  ACM, \bibinfo{pages}{349--360}.
\newblock


\bibitem[\protect\citeauthoryear{Efron}{Efron}{1982}]%
        {efron1982jackknife}
\bibfield{author}{\bibinfo{person}{Bradley Efron}.}
  \bibinfo{year}{1982}\natexlab{}.
\newblock \bibinfo{booktitle}{{\em The jackknife, the bootstrap and other
  resampling plans}}.
\newblock \bibinfo{publisher}{SIAM}.
\newblock


\bibitem[\protect\citeauthoryear{Feild, Binkley, and Lawrie}{Feild
  et~al\mbox{.}}{2006}]%
        {feild2006empirical}
\bibfield{author}{\bibinfo{person}{Henry Feild}, \bibinfo{person}{David
  Binkley}, {and} \bibinfo{person}{Dawn Lawrie}.}
  \bibinfo{year}{2006}\natexlab{}.
\newblock \showarticletitle{An empirical comparison of techniques for
  extracting concept abbreviations from identifiers}. In
  \bibinfo{booktitle}{{\em Proceedings of IASTED International Conference on
  Software Engineering and Applications (SEA’06)}}. Citeseer.
\newblock


\bibitem[\protect\citeauthoryear{Feng, Martins, Geffen, Dillig, and
  Chaudhuri}{Feng et~al\mbox{.}}{2016}]%
        {feng2016}
\bibfield{author}{\bibinfo{person}{Yu Feng}, \bibinfo{person}{Ruben Martins},
  \bibinfo{person}{Jacob~Van Geffen}, \bibinfo{person}{Isil Dillig}, {and}
  \bibinfo{person}{Swarat Chaudhuri}.} \bibinfo{year}{2016}\natexlab{}.
\newblock \showarticletitle{Component-based Synthesis of Table Consolidation
  and Transformation Tasks from Examples}.
\newblock \bibinfo{journal}{{\em CoRR\/}}  \bibinfo{volume}{abs/1611.07502}
  (\bibinfo{year}{2016}).
\newblock
\showURL{%
\url{http://arxiv.org/abs/1611.07502}}


\bibitem[\protect\citeauthoryear{Feng, Martins, Wang, Dillig, and Reps}{Feng
  et~al\mbox{.}}{2017}]%
        {feng2017}
\bibfield{author}{\bibinfo{person}{Yu Feng}, \bibinfo{person}{Ruben Martins},
  \bibinfo{person}{Yuepeng Wang}, \bibinfo{person}{Isil Dillig}, {and}
  \bibinfo{person}{Thomas~W. Reps}.} \bibinfo{year}{2017}\natexlab{}.
\newblock \showarticletitle{Component-based Synthesis for Complex APIs}. In
  \bibinfo{booktitle}{{\em Proceedings of the 44th ACM SIGPLAN Symposium on
  Principles of Programming Languages}} {\em (\bibinfo{series}{POPL 2017})}.
  \bibinfo{publisher}{ACM}, \bibinfo{address}{New York, NY, USA},
  \bibinfo{pages}{599--612}.
\newblock
\showISBNx{978-1-4503-4660-3}
\showDOI{%
\url{http://dx.doi.org/10.1145/3009837.3009851}}


\bibitem[\protect\citeauthoryear{Feser, Chaudhuri, and Dillig}{Feser
  et~al\mbox{.}}{2015}]%
        {Feser2015}
\bibfield{author}{\bibinfo{person}{John~K. Feser}, \bibinfo{person}{Swarat
  Chaudhuri}, {and} \bibinfo{person}{Isil Dillig}.}
  \bibinfo{year}{2015}\natexlab{}.
\newblock \showarticletitle{Synthesizing Data Structure Transformations from
  Input-output Examples}. In \bibinfo{booktitle}{{\em Proceedings of the 36th
  ACM SIGPLAN Conference on Programming Language Design and Implementation}}
  {\em (\bibinfo{series}{PLDI 2015})}. \bibinfo{publisher}{ACM},
  \bibinfo{address}{New York, NY, USA}, \bibinfo{pages}{229--239}.
\newblock
\showISBNx{978-1-4503-3468-6}
\showDOI{%
\url{http://dx.doi.org/10.1145/2737924.2737977}}


\bibitem[\protect\citeauthoryear{Grechanik, Fu, Xie, McMillan, Poshyvanyk, and
  Cumby}{Grechanik et~al\mbox{.}}{2010}]%
        {Grechanik2010}
\bibfield{author}{\bibinfo{person}{Mark Grechanik}, \bibinfo{person}{Chen Fu},
  \bibinfo{person}{Qing Xie}, \bibinfo{person}{Collin McMillan},
  \bibinfo{person}{Denys Poshyvanyk}, {and} \bibinfo{person}{Chad Cumby}.}
  \bibinfo{year}{2010}\natexlab{}.
\newblock \showarticletitle{{A search engine for finding highly relevant
  applications}}. In \bibinfo{booktitle}{{\em ACM/IEEE International Conference
  on Software Engineering}}. \bibinfo{publisher}{ACM Press},
  \bibinfo{address}{New York, New York, USA}.
\newblock


\bibitem[\protect\citeauthoryear{Gvero and Kuncak}{Gvero and Kuncak}{2015}]%
        {gvero2015interactive}
\bibfield{author}{\bibinfo{person}{Tihomir Gvero} {and} \bibinfo{person}{Viktor
  Kuncak}.} \bibinfo{year}{2015}\natexlab{}.
\newblock \showarticletitle{Interactive synthesis using free-form queries}. In
  \bibinfo{booktitle}{{\em 2015 IEEE/ACM 37th IEEE International Conference on
  Software Engineering}}, Vol.~\bibinfo{volume}{2}. IEEE,
  \bibinfo{pages}{689--692}.
\newblock


\bibitem[\protect\citeauthoryear{Harman, Jia, and Langdon}{Harman
  et~al\mbox{.}}{2014}]%
        {harman2014babel}
\bibfield{author}{\bibinfo{person}{Mark Harman}, \bibinfo{person}{Yue Jia},
  {and} \bibinfo{person}{William~B Langdon}.} \bibinfo{year}{2014}\natexlab{}.
\newblock \showarticletitle{Babel pidgin: SBSE can grow and graft entirely new
  functionality into a real world system}. In \bibinfo{booktitle}{{\em
  International Symposium on Search Based Software Engineering}}. Springer,
  \bibinfo{pages}{247--252}.
\newblock


\bibitem[\protect\citeauthoryear{Hindle, Barr, Su, Gabel, and Devanbu}{Hindle
  et~al\mbox{.}}{2012}]%
        {hindle2012naturalness}
\bibfield{author}{\bibinfo{person}{Abram Hindle}, \bibinfo{person}{Earl~T
  Barr}, \bibinfo{person}{Zhendong Su}, \bibinfo{person}{Mark Gabel}, {and}
  \bibinfo{person}{Premkumar Devanbu}.} \bibinfo{year}{2012}\natexlab{}.
\newblock \showarticletitle{On the naturalness of software}. In
  \bibinfo{booktitle}{{\em 2012 34th International Conference on Software
  Engineering (ICSE)}}. IEEE, \bibinfo{pages}{837--847}.
\newblock


\bibitem[\protect\citeauthoryear{Holmes and Walker}{Holmes and Walker}{2013}]%
        {Holmes2013}
\bibfield{author}{\bibinfo{person}{Reid Holmes} {and}
  \bibinfo{person}{Robert~J. Walker}.} \bibinfo{year}{2013}\natexlab{}.
\newblock \showarticletitle{Systematizing Pragmatic Software Reuse}.
\newblock \bibinfo{journal}{{\em ACM Trans. Softw. Eng. Methodol.\/}}
  \bibinfo{volume}{21}, \bibinfo{number}{4}, Article \bibinfo{articleno}{20}
  (\bibinfo{date}{Feb.} \bibinfo{year}{2013}), \bibinfo{numpages}{44}~pages.
\newblock
\showISSN{1049-331X}
\showDOI{%
\url{http://dx.doi.org/10.1145/2377656.2377657}}


\bibitem[\protect\citeauthoryear{Hummel, Janjic, and Atkinson}{Hummel
  et~al\mbox{.}}{2008}]%
        {hummel2008code}
\bibfield{author}{\bibinfo{person}{Oliver Hummel}, \bibinfo{person}{Werner
  Janjic}, {and} \bibinfo{person}{Colin Atkinson}.}
  \bibinfo{year}{2008}\natexlab{}.
\newblock \showarticletitle{Code conjurer: Pulling reusable software out of
  thin air}.
\newblock \bibinfo{journal}{{\em IEEE software\/}} \bibinfo{volume}{25},
  \bibinfo{number}{5} (\bibinfo{year}{2008}).
\newblock


\bibitem[\protect\citeauthoryear{Jia, Harman, Langdon, and Marginean}{Jia
  et~al\mbox{.}}{2015}]%
        {jia2015grow}
\bibfield{author}{\bibinfo{person}{Yue Jia}, \bibinfo{person}{Mark Harman},
  \bibinfo{person}{William~B Langdon}, {and} \bibinfo{person}{Alexandru
  Marginean}.} \bibinfo{year}{2015}\natexlab{}.
\newblock \showarticletitle{Grow and serve: Growing Django citation services
  using SBSE}. In \bibinfo{booktitle}{{\em International Symposium on Search
  Based Software Engineering}}. Springer, \bibinfo{pages}{269--275}.
\newblock


\bibitem[\protect\citeauthoryear{Jiang, Misherghi, Su, and Glondu}{Jiang
  et~al\mbox{.}}{2007}]%
        {jiang2007deckard}
\bibfield{author}{\bibinfo{person}{Lingxiao Jiang}, \bibinfo{person}{Ghassan
  Misherghi}, \bibinfo{person}{Zhendong Su}, {and} \bibinfo{person}{Stephane
  Glondu}.} \bibinfo{year}{2007}\natexlab{}.
\newblock \showarticletitle{Deckard: Scalable and accurate tree-based detection
  of code clones}. In \bibinfo{booktitle}{{\em Proceedings of the 29th
  international conference on Software Engineering}}. IEEE Computer Society,
  \bibinfo{pages}{96--105}.
\newblock


\bibitem[\protect\citeauthoryear{Juergens, Deissenboeck, Hummel, and
  Wagner}{Juergens et~al\mbox{.}}{2009}]%
        {juergens2009code}
\bibfield{author}{\bibinfo{person}{Elmar Juergens}, \bibinfo{person}{Florian
  Deissenboeck}, \bibinfo{person}{Benjamin Hummel}, {and}
  \bibinfo{person}{Stefan Wagner}.} \bibinfo{year}{2009}\natexlab{}.
\newblock \showarticletitle{Do code clones matter?}. In
  \bibinfo{booktitle}{{\em Proceedings of the 31st International Conference on
  Software Engineering}}. IEEE Computer Society, \bibinfo{pages}{485--495}.
\newblock


\bibitem[\protect\citeauthoryear{Ke, Stolee, Le~Goues, and Brun}{Ke
  et~al\mbox{.}}{2015}]%
        {ke2015repairing}
\bibfield{author}{\bibinfo{person}{Yalin Ke}, \bibinfo{person}{Kathryn~T
  Stolee}, \bibinfo{person}{Claire Le~Goues}, {and} \bibinfo{person}{Yuriy
  Brun}.} \bibinfo{year}{2015}\natexlab{}.
\newblock \showarticletitle{Repairing programs with semantic code search (t)}.
  In \bibinfo{booktitle}{{\em Automated Software Engineering (ASE), 2015 30th
  IEEE/ACM International Conference on}}. IEEE, \bibinfo{pages}{295--306}.
\newblock


\bibitem[\protect\citeauthoryear{Keivanloo, Rilling, and Zou}{Keivanloo
  et~al\mbox{.}}{2014}]%
        {keivanloo2014spotting}
\bibfield{author}{\bibinfo{person}{Iman Keivanloo}, \bibinfo{person}{Juergen
  Rilling}, {and} \bibinfo{person}{Ying Zou}.} \bibinfo{year}{2014}\natexlab{}.
\newblock \showarticletitle{Spotting working code examples}. In
  \bibinfo{booktitle}{{\em Proceedings of the 36th International Conference on
  Software Engineering}}. ACM, \bibinfo{pages}{664--675}.
\newblock


\bibitem[\protect\citeauthoryear{Kim, Bergman, Lau, and Notkin}{Kim
  et~al\mbox{.}}{2004}]%
        {kim2004ethnographic}
\bibfield{author}{\bibinfo{person}{Miryung Kim}, \bibinfo{person}{Lawrence
  Bergman}, \bibinfo{person}{Tessa Lau}, {and} \bibinfo{person}{David Notkin}.}
  \bibinfo{year}{2004}\natexlab{}.
\newblock \showarticletitle{An ethnographic study of copy and paste programming
  practices in OOPL}. In \bibinfo{booktitle}{{\em Empirical Software
  Engineering, 2004. ISESE'04. Proceedings. 2004 International Symposium on}}.
  IEEE, \bibinfo{pages}{83--92}.
\newblock


\bibitem[\protect\citeauthoryear{Lazzarini~Lemos, Bajracharya, Ossher, Masiero,
  and Lopes}{Lazzarini~Lemos et~al\mbox{.}}{2009}]%
        {lazzarini2009applying}
\bibfield{author}{\bibinfo{person}{Ot{\'a}vio~Augusto Lazzarini~Lemos},
  \bibinfo{person}{Sushil Bajracharya}, \bibinfo{person}{Joel Ossher},
  \bibinfo{person}{Paulo~Cesar Masiero}, {and} \bibinfo{person}{Cristina
  Lopes}.} \bibinfo{year}{2009}\natexlab{}.
\newblock \showarticletitle{Applying test-driven code search to the reuse of
  auxiliary functionality}. In \bibinfo{booktitle}{{\em Proceedings of the 2009
  ACM symposium on Applied Computing}}. ACM, \bibinfo{pages}{476--482}.
\newblock


\bibitem[\protect\citeauthoryear{Marginean, Barr, Harman, and Jia}{Marginean
  et~al\mbox{.}}{2015}]%
        {marginean2015automated}
\bibfield{author}{\bibinfo{person}{Alexandru Marginean},
  \bibinfo{person}{Earl~T Barr}, \bibinfo{person}{Mark Harman}, {and}
  \bibinfo{person}{Yue Jia}.} \bibinfo{year}{2015}\natexlab{}.
\newblock \showarticletitle{Automated transplantation of call graph and layout
  features into Kate}. In \bibinfo{booktitle}{{\em International Symposium on
  Search Based Software Engineering}}. Springer, \bibinfo{pages}{262--268}.
\newblock


\bibitem[\protect\citeauthoryear{McMillan, Grechanik, Poshyvanyk, Xie, and
  Fu}{McMillan et~al\mbox{.}}{2011}]%
        {McMillan2011}
\bibfield{author}{\bibinfo{person}{Collin McMillan}, \bibinfo{person}{Mark
  Grechanik}, \bibinfo{person}{Denys Poshyvanyk}, \bibinfo{person}{Qing Xie},
  {and} \bibinfo{person}{Chen Fu}.} \bibinfo{year}{2011}\natexlab{}.
\newblock \showarticletitle{{Portfolio: finding relevant functions and their
  usage}}. In \bibinfo{booktitle}{{\em International conference on Software
  engineering}}. \bibinfo{publisher}{ACM Press}, \bibinfo{address}{New York,
  New York, USA}.
\newblock


\bibitem[\protect\citeauthoryear{Mishne, Shoham, and Yahav}{Mishne
  et~al\mbox{.}}{2012}]%
        {mishne12}
\bibfield{author}{\bibinfo{person}{Alon Mishne}, \bibinfo{person}{Sharon
  Shoham}, {and} \bibinfo{person}{Eran Yahav}.}
  \bibinfo{year}{2012}\natexlab{}.
\newblock \showarticletitle{{Typestate-based Semantic Code Search over Partial
  Programs}}. In \bibinfo{booktitle}{{\em Proceedings of the ACM International
  Conference on Object Oriented Programming Systems Languages and
  Applications}} {\em (\bibinfo{series}{OOPSLA '12})}.
  \bibinfo{publisher}{ACM}, \bibinfo{address}{New York, NY, USA},
  \bibinfo{pages}{997--1016}.
\newblock
\showISBNx{978-1-4503-1561-6}
\showDOI{%
\url{http://dx.doi.org/10.1145/2384616.2384689}}


\bibitem[\protect\citeauthoryear{Narasimhan and Reichenbach}{Narasimhan and
  Reichenbach}{2015}]%
        {narasimhan2015copy}
\bibfield{author}{\bibinfo{person}{Krishna Narasimhan} {and}
  \bibinfo{person}{Christoph Reichenbach}.} \bibinfo{year}{2015}\natexlab{}.
\newblock \showarticletitle{Copy and Paste Redeemed}. In
  \bibinfo{booktitle}{{\em Automated Software Engineering (ASE), 2015 30th
  IEEE/ACM International Conference on}}. IEEE, \bibinfo{pages}{630--640}.
\newblock


\bibitem[\protect\citeauthoryear{Nguyen and Nguyen}{Nguyen and Nguyen}{2015}]%
        {nguyen2015graph}
\bibfield{author}{\bibinfo{person}{Anh~Tuan Nguyen} {and}
  \bibinfo{person}{Tien~N Nguyen}.} \bibinfo{year}{2015}\natexlab{}.
\newblock \showarticletitle{Graph-based statistical language model for code}.
  In \bibinfo{booktitle}{{\em Proceedings of the 37th International Conference
  on Software Engineering-Volume 1}}. IEEE Press, \bibinfo{pages}{858--868}.
\newblock


\bibitem[\protect\citeauthoryear{Ossher, Sajnani, and Lopes}{Ossher
  et~al\mbox{.}}{2012}]%
        {Ossher:WCRE12}
\bibfield{author}{\bibinfo{person}{J. Ossher}, \bibinfo{person}{H. Sajnani},
  {and} \bibinfo{person}{C. Lopes}.} \bibinfo{year}{2012}\natexlab{}.
\newblock \showarticletitle{Astra: Bottom-up Construction of Structured
  Artifact Repositories}. In \bibinfo{booktitle}{{\em Reverse Engineering
  (WCRE), 2012 19th Working Conference on}}. \bibinfo{pages}{41--50}.
\newblock
\showISSN{1095-1350}
\showDOI{%
\url{http://dx.doi.org/10.1109/WCRE.2012.14}}


\bibitem[\protect\citeauthoryear{Petke, Harman, Langdon, and Weimer}{Petke
  et~al\mbox{.}}{2014}]%
        {petke2014using}
\bibfield{author}{\bibinfo{person}{Justyna Petke}, \bibinfo{person}{Mark
  Harman}, \bibinfo{person}{William~B Langdon}, {and} \bibinfo{person}{Westley
  Weimer}.} \bibinfo{year}{2014}\natexlab{}.
\newblock \showarticletitle{Using genetic improvement and code transplants to
  specialise a C++ program to a problem class}. In \bibinfo{booktitle}{{\em
  European Conference on Genetic Programming}}. Springer,
  \bibinfo{pages}{137--149}.
\newblock


\bibitem[\protect\citeauthoryear{Pnueli and Rosner}{Pnueli and Rosner}{1989}]%
        {Pnueli1989}
\bibfield{author}{\bibinfo{person}{Amir Pnueli} {and} \bibinfo{person}{Roni
  Rosner}.} \bibinfo{year}{1989}\natexlab{}.
\newblock \showarticletitle{On the Synthesis of an Asynchronous Reactive
  Module}. In \bibinfo{booktitle}{{\em Proceedings of the 16th International
  Colloquium on Automata, Languages and Programming}} {\em
  (\bibinfo{series}{ICALP '89})}. \bibinfo{publisher}{Springer-Verlag},
  \bibinfo{address}{London, UK, UK}, \bibinfo{pages}{652--671}.
\newblock
\showISBNx{3-540-51371-X}
\showURL{%
\url{http://dl.acm.org/citation.cfm?id=646243.681607}}


\bibitem[\protect\citeauthoryear{Polozov and Gulwani}{Polozov and
  Gulwani}{2015}]%
        {polozov2015flashmeta}
\bibfield{author}{\bibinfo{person}{Oleksandr Polozov} {and}
  \bibinfo{person}{Sumit Gulwani}.} \bibinfo{year}{2015}\natexlab{}.
\newblock \showarticletitle{Flashmeta: A framework for inductive program
  synthesis}.
\newblock \bibinfo{journal}{{\em ACM SIGPLAN Notices\/}} \bibinfo{volume}{50},
  \bibinfo{number}{10} (\bibinfo{year}{2015}), \bibinfo{pages}{107--126}.
\newblock


\bibitem[\protect\citeauthoryear{Raghothaman, Wei, and Hamadi}{Raghothaman
  et~al\mbox{.}}{2015}]%
        {raghothaman2015swim}
\bibfield{author}{\bibinfo{person}{Mukund Raghothaman}, \bibinfo{person}{Yi
  Wei}, {and} \bibinfo{person}{Youssef Hamadi}.}
  \bibinfo{year}{2015}\natexlab{}.
\newblock \showarticletitle{SWIM: Synthesizing What I Mean}.
\newblock \bibinfo{journal}{{\em arXiv preprint arXiv:1511.08497\/}}
  (\bibinfo{year}{2015}).
\newblock


\bibitem[\protect\citeauthoryear{Raychev, Bielik, Vechev, and Krause}{Raychev
  et~al\mbox{.}}{2016}]%
        {raychev2016}
\bibfield{author}{\bibinfo{person}{Veselin Raychev}, \bibinfo{person}{Pavol
  Bielik}, \bibinfo{person}{Martin Vechev}, {and} \bibinfo{person}{Andreas
  Krause}.} \bibinfo{year}{2016}\natexlab{}.
\newblock \showarticletitle{Learning programs from noisy data}. In
  \bibinfo{booktitle}{{\em ACM SIGPLAN Notices}}, Vol.~\bibinfo{volume}{51}.
  ACM, \bibinfo{pages}{761--774}.
\newblock


\bibitem[\protect\citeauthoryear{Raychev, Vechev, and Krause}{Raychev
  et~al\mbox{.}}{2015}]%
        {Raychev2015}
\bibfield{author}{\bibinfo{person}{Veselin Raychev}, \bibinfo{person}{Martin
  Vechev}, {and} \bibinfo{person}{Andreas Krause}.}
  \bibinfo{year}{2015}\natexlab{}.
\newblock \showarticletitle{Predicting Program Properties from "Big Code"}. In
  \bibinfo{booktitle}{{\em Proceedings of the 42Nd Annual ACM SIGPLAN-SIGACT
  Symposium on Principles of Programming Languages}} {\em
  (\bibinfo{series}{POPL '15})}. \bibinfo{publisher}{ACM},
  \bibinfo{address}{New York, NY, USA}, \bibinfo{pages}{111--124}.
\newblock
\showISBNx{978-1-4503-3300-9}
\showDOI{%
\url{http://dx.doi.org/10.1145/2676726.2677009}}


\bibitem[\protect\citeauthoryear{Raychev, Vechev, and Yahav}{Raychev
  et~al\mbox{.}}{2014}]%
        {Raychev2014}
\bibfield{author}{\bibinfo{person}{Veselin Raychev}, \bibinfo{person}{Martin
  Vechev}, {and} \bibinfo{person}{Eran Yahav}.}
  \bibinfo{year}{2014}\natexlab{}.
\newblock \showarticletitle{Code Completion with Statistical Language Models}.
  In \bibinfo{booktitle}{{\em Proceedings of the 35th ACM SIGPLAN Conference on
  Programming Language Design and Implementation}} {\em (\bibinfo{series}{PLDI
  '14})}. \bibinfo{publisher}{ACM}, \bibinfo{address}{New York, NY, USA},
  \bibinfo{pages}{419--428}.
\newblock
\showISBNx{978-1-4503-2784-8}
\showDOI{%
\url{http://dx.doi.org/10.1145/2594291.2594321}}


\bibitem[\protect\citeauthoryear{Reiss}{Reiss}{2009}]%
        {reiss2009}
\bibfield{author}{\bibinfo{person}{Steven~P. Reiss}.}
  \bibinfo{year}{2009}\natexlab{}.
\newblock \showarticletitle{Semantics-based Code Search}. In
  \bibinfo{booktitle}{{\em Proceedings of the 31st International Conference on
  Software Engineering}} {\em (\bibinfo{series}{ICSE '09})}.
  \bibinfo{publisher}{IEEE Computer Society}, \bibinfo{address}{Washington, DC,
  USA}, \bibinfo{pages}{243--253}.
\newblock
\showISBNx{978-1-4244-3453-4}
\showDOI{%
\url{http://dx.doi.org/10.1109/ICSE.2009.5070525}}


\bibitem[\protect\citeauthoryear{Roy, Cordy, and Koschke}{Roy
  et~al\mbox{.}}{2009}]%
        {roy2009comparison}
\bibfield{author}{\bibinfo{person}{Chanchal~K Roy}, \bibinfo{person}{James~R
  Cordy}, {and} \bibinfo{person}{Rainer Koschke}.}
  \bibinfo{year}{2009}\natexlab{}.
\newblock \showarticletitle{Comparison and evaluation of code clone detection
  techniques and tools: A qualitative approach}.
\newblock \bibinfo{journal}{{\em Science of Computer Programming\/}}
  \bibinfo{volume}{74}, \bibinfo{number}{7} (\bibinfo{year}{2009}),
  \bibinfo{pages}{470--495}.
\newblock


\bibitem[\protect\citeauthoryear{Sajnani, Saini, Ossher, and Lopes}{Sajnani
  et~al\mbox{.}}{2014}]%
        {sajnani:icsem2014}
\bibfield{author}{\bibinfo{person}{H. Sajnani}, \bibinfo{person}{V. Saini},
  \bibinfo{person}{J. Ossher}, {and} \bibinfo{person}{C.V. Lopes}.}
  \bibinfo{year}{2014}\natexlab{}.
\newblock \showarticletitle{Is Popularity a Measure of Quality? An Analysis of
  Maven Components}. In \bibinfo{booktitle}{{\em Software Maintenance and
  Evolution (ICSME), 2014 IEEE International Conference on}}.
  \bibinfo{pages}{231--240}.
\newblock
\showISSN{1063-6773}
\showDOI{%
\url{http://dx.doi.org/10.1109/ICSME.2014.45}}


\bibitem[\protect\citeauthoryear{Sidiroglou-Douskos, Lahtinen, Long, and
  Rinard}{Sidiroglou-Douskos et~al\mbox{.}}{2015}]%
        {sidiroglou2015automatic}
\bibfield{author}{\bibinfo{person}{Stelios Sidiroglou-Douskos},
  \bibinfo{person}{Eric Lahtinen}, \bibinfo{person}{Fan Long}, {and}
  \bibinfo{person}{Martin Rinard}.} \bibinfo{year}{2015}\natexlab{}.
\newblock \showarticletitle{Automatic error elimination by horizontal code
  transfer across multiple applications}. In \bibinfo{booktitle}{{\em ACM
  SIGPLAN Notices}}, Vol.~\bibinfo{volume}{50}. ACM, \bibinfo{pages}{43--54}.
\newblock


\bibitem[\protect\citeauthoryear{Solar-Lezama}{Solar-Lezama}{2009}]%
        {solar2009sketching}
\bibfield{author}{\bibinfo{person}{Armando Solar-Lezama}.}
  \bibinfo{year}{2009}\natexlab{}.
\newblock \showarticletitle{The sketching approach to program synthesis}. In
  \bibinfo{booktitle}{{\em Asian Symposium on Programming Languages and
  Systems}}. Springer, \bibinfo{pages}{4--13}.
\newblock


\bibitem[\protect\citeauthoryear{Solar-Lezama, Tancau, Bodik, Seshia, and
  Saraswat}{Solar-Lezama et~al\mbox{.}}{2006}]%
        {lezama06}
\bibfield{author}{\bibinfo{person}{Armando Solar-Lezama},
  \bibinfo{person}{Liviu Tancau}, \bibinfo{person}{Rastislav Bodik},
  \bibinfo{person}{Sanjit Seshia}, {and} \bibinfo{person}{Vijay Saraswat}.}
  \bibinfo{year}{2006}\natexlab{}.
\newblock \showarticletitle{Combinatorial Sketching for Finite Programs}. In
  \bibinfo{booktitle}{{\em Proceedings of the 12th International Conference on
  Architectural Support for Programming Languages and Operating Systems}} {\em
  (\bibinfo{series}{ASPLOS XII})}. \bibinfo{publisher}{ACM},
  \bibinfo{address}{New York, NY, USA}, \bibinfo{pages}{404--415}.
\newblock
\showISBNx{1-59593-451-0}
\showDOI{%
\url{http://dx.doi.org/10.1145/1168857.1168907}}


\bibitem[\protect\citeauthoryear{Srivastava, Gulwani, and Foster}{Srivastava
  et~al\mbox{.}}{2012}]%
        {Srivastava2012}
\bibfield{author}{\bibinfo{person}{Saurabh Srivastava}, \bibinfo{person}{Sumit
  Gulwani}, {and} \bibinfo{person}{Jeffrey~S. Foster}.}
  \bibinfo{year}{2012}\natexlab{}.
\newblock \showarticletitle{Template-based program verification and program
  synthesis}.
\newblock \bibinfo{journal}{{\em International Journal on Software Tools for
  Technology Transfer\/}} \bibinfo{volume}{15}, \bibinfo{number}{5}
  (\bibinfo{year}{2012}), \bibinfo{pages}{497--518}.
\newblock
\showISSN{1433-2787}
\showDOI{%
\url{http://dx.doi.org/10.1007/s10009-012-0223-4}}


\bibitem[\protect\citeauthoryear{Stolee and Elbaum}{Stolee and Elbaum}{2012}]%
        {stolee2012toward}
\bibfield{author}{\bibinfo{person}{Kathryn~T Stolee} {and}
  \bibinfo{person}{Sebastian Elbaum}.} \bibinfo{year}{2012}\natexlab{}.
\newblock \showarticletitle{Toward semantic search via SMT solver}. In
  \bibinfo{booktitle}{{\em Proceedings of the ACM SIGSOFT 20th International
  Symposium on the Foundations of Software Engineering}}. ACM,
  \bibinfo{pages}{25}.
\newblock


\bibitem[\protect\citeauthoryear{Udupa, Raghavan, Deshmukh, Mador-Haim, Martin,
  and Alur}{Udupa et~al\mbox{.}}{2013}]%
        {udupa2013transit}
\bibfield{author}{\bibinfo{person}{Abhishek Udupa}, \bibinfo{person}{Arun
  Raghavan}, \bibinfo{person}{Jyotirmoy~V Deshmukh}, \bibinfo{person}{Sela
  Mador-Haim}, \bibinfo{person}{Milo~MK Martin}, {and} \bibinfo{person}{Rajeev
  Alur}.} \bibinfo{year}{2013}\natexlab{}.
\newblock \showarticletitle{{TRANSIT}: specifying protocols with concolic
  snippets}.
\newblock \bibinfo{journal}{{\em ACM SIGPLAN Notices\/}} \bibinfo{volume}{48},
  \bibinfo{number}{6} (\bibinfo{year}{2013}), \bibinfo{pages}{287--296}.
\newblock


\bibitem[\protect\citeauthoryear{Wang, Feng, Martins, Kaushik, Dillig, and
  Reiss}{Wang et~al\mbox{.}}{2016}]%
        {wang2016}
\bibfield{author}{\bibinfo{person}{Yuepeng Wang}, \bibinfo{person}{Yu Feng},
  \bibinfo{person}{Ruben Martins}, \bibinfo{person}{Arati Kaushik},
  \bibinfo{person}{Isil Dillig}, {and} \bibinfo{person}{Steven~P. Reiss}.}
  \bibinfo{year}{2016}\natexlab{}.
\newblock \showarticletitle{Hunter: Next-generation Code Reuse for Java}. In
  \bibinfo{booktitle}{{\em Proceedings of the 2016 24th ACM SIGSOFT
  International Symposium on Foundations of Software Engineering}} {\em
  (\bibinfo{series}{FSE 2016})}. \bibinfo{publisher}{ACM},
  \bibinfo{address}{New York, NY, USA}, \bibinfo{pages}{1028--1032}.
\newblock
\showISBNx{978-1-4503-4218-6}
\showDOI{%
\url{http://dx.doi.org/10.1145/2950290.2983934}}


\bibitem[\protect\citeauthoryear{Yaghmazadeh, Klinger, Dillig, and
  Chaudhuri}{Yaghmazadeh et~al\mbox{.}}{2016}]%
        {yaghmazadeh2016}
\bibfield{author}{\bibinfo{person}{Navid Yaghmazadeh},
  \bibinfo{person}{Christian Klinger}, \bibinfo{person}{Isil Dillig}, {and}
  \bibinfo{person}{Swarat Chaudhuri}.} \bibinfo{year}{2016}\natexlab{}.
\newblock \showarticletitle{Synthesizing Transformations on Hierarchically
  Structured Data}. In \bibinfo{booktitle}{{\em Proceedings of the 37th ACM
  SIGPLAN Conference on Programming Language Design and Implementation}} {\em
  (\bibinfo{series}{PLDI '16})}. \bibinfo{publisher}{ACM},
  \bibinfo{address}{New York, NY, USA}, \bibinfo{pages}{508--521}.
\newblock
\showISBNx{978-1-4503-4261-2}
\showDOI{%
\url{http://dx.doi.org/10.1145/2908080.2908088}}


\end{thebibliography}

\appendix
\section{Appendix}
\subsection{Draft Programs in Benchmarks}
\label{subsec:draft}
\begin{description}
\item[Binary Search Draft] \hfill \\
\begin{lstlisting}
public class Binsearch {

  /**
   * COMMENT:
   * Binary search
   *
   * TEST:
   * int [] array = new int [] {-100, -20, 0, 4, 100, 600, 601};
   * __solution__
   * return(
   * binsearch(array, -100) == 0 &&
   * binsearch(array, 100) == 4 &&
   * binsearch(array, 602) == -1);
   */
  public int binsearch(int[] array, int x) {
    int result = -1;
    int low = 0;
    int high = ??;

    ??
    return result;
  }
}
\end{lstlisting}
  \newpage
\item[Collision Detection Draft] \hfill \\
\begin{lstlisting}
import net.smert.jreactphysics3d.mathematics.Vector3;
import net.smert.jreactphysics3d.collision.shapes.AABB;
public class AABB2 {
  /**
   * COMMENT:
   * Testing whether two AABB objects collide
   *
   * TEST:
   * public class Vector3 { ... }
   *
   * public class AABB { ... }
   *
   * __solution__
   * public boolean test_no_collision() {
   *   Vector3 a_min = new Vector3(0, 0, 0);
   *   Vector3 a_max = new Vector3(1, 1, 1);
   *   AABB a = new AABB(a_min, a_max);
   *
   *   Vector3 b_min = new Vector3(10, 10, 10);
   *   Vector3 b_max = new Vector3(20, 20, 20);
   *   AABB b = new AABB(b_min, b_max);
   *
   *   if(testCollision(a, b)) {
   *     return false;
   *   }
   *
   *   Vector3 c_min = new Vector3(0, 0, 0);
   *   Vector3 c_max = new Vector3(10, 10, 10);
   *   AABB c = new AABB(c_min, c_max);
   *
   *   Vector3 d_min = new Vector3(0, 0, 30);
   *   Vector3 d_max = new Vector3(20, 20, 2);
   *   AABB d = new AABB(d_min, d_max);
   *
   *   if(testCollision(c, d)) {
   *     return false;
   *   }
   *
   *   Vector3 e_min = new Vector3(0, 0, 0);
   *   Vector3 e_max = new Vector3(10, 10, 10);
   *   AABB e = new AABB(e_min, e_max);
   *
   *   Vector3 f_min = new Vector3(30, 0, 0);
   *   Vector3 f_max = new Vector3(50, 20, 2);
   *   AABB f = new AABB(f_min, f_max);
   *
   *   if(testCollision(e, f)) {
   *   return false;
   *   }
   *
   *   return true;
   * }
   *
   * public boolean test_collision() {
   *   Vector3 a_min = new Vector3(0, 0, 0);
   *   Vector3 a_max = new Vector3(10, 10, 10);
   *   AABB a = new AABB(a_min, a_max);
   *
   *   Vector3 b_min = new Vector3(0, 0, 0);
   *   Vector3 b_max = new Vector3(20, 20, 2);
   *   AABB b = new AABB(b_min, b_max);
   *
   *   if(!testCollision(a, b)) {
   *     return false;
   *   }
   *
   *   return true;
   * }
   *
   * public void test() {
   *   if(!test_no_collision() || !test_collision()) {
   *     return("false");
   *   }
   *   return("true");
   * }
   * test();
   *
   */
    public boolean testCollision(AABB a, AABB aabb) {
      ??
      return ?? && ??;
    }

}
\end{lstlisting}
  \newpage
\item[CSV Draft] \hfill \\
\begin{lstlisting}
import java.io.File;
import java.util.Scanner;
public class CSV {
    /**
     * COMMENT:
     * Reading a matrix from a CSV file
     *
     * TEST:
     * import java.io.File;
     * import java.util.*;
     * __solution__
     * int num_elements = 0;
     * int[][] mat = read_csv("data.csv");
     * for(int i = 0; i < mat.length; ++i) {
     *   for(int j = 0; j < mat[0].length; ++j) {
     *     if(i + j + 1 != mat[i][j]) {
     *       return("false");
     *       exit();
     *     }
     *     num_elements += 1;
     *   }
     * }
     * return(num_elements == 20);
     */
    public int[][] read_csv(String filename) {
        int row, col;
        File f = new File(??);
        /**
         * COMMENT:
         * Reading a matrix from a CSV file
         *
         * TEST:
         * import java.io.File;
         * import java.util.*;
         * int row, col;
         * File f = new File("data.csv");
         * __solution__
         * return(row == 5 && col == 4);
         */
        ??

        int[][] mat = new int[row][col];

        // matrix multiplication
        ??
        return mat;
    }
}
\end{lstlisting}
  \newpage
\item[Echo Server Draft] \hfill \\
\begin{lstlisting}
public class Echo {
    /**
     * COMMENT:
     * Echo server in java
     *
     * TEST:
     * API_cons("echo-server-test.java");
     * __solution__
     * run();
     * return(
     *   has_server_sock &&
     *   has_client_sock &&
     *   has_in &&
     *   has_out &&
     *   has_readline &&
     *   has_println &&
     *   in_closed &&
     *   out_closed &&
     *   client_closed);
     */
    public void run() {
        // create socket
        int port = 4444;
        SServerSocket serverSocket = new SServerSocket(??);
        ??
    }
}
\end{lstlisting}
  \newpage
\item[Face Detection Draft] \hfill \\
\begin{lstlisting}
import org.opencv.core.Core;
import org.opencv.core.Mat;
import org.opencv.core.MatOfRect;
import org.opencv.core.Point;
import org.opencv.core.Rect;
import org.opencv.core.Scalar;
import org.opencv.imgcodecs.Imgcodecs;
import org.opencv.imgproc.Imgproc;
import org.opencv.highgui.Highgui;
import org.opencv.objdetect.CascadeClassifier;

public class DetectFaceDemo {
    /**
     * COMMENT:
     * Doing face detection using OpenCV
     *
     * TEST:
     * API_cons("face-detection-test.java");
     * __solution__
     * run();
     * return(_has_detector_ && _has_image_ && _has_detection_ && _image_written_);
     *
     */
    public void run() {
        String input_img = "lena.jpg";
        String output_img = "faceDetection.png";
    	CascadeClassifier faceDetector = new CascadeClassifier(??);
        ??
    }
}
\end{lstlisting}
  \newpage
\item[Collecting Files Draft] \hfill \\
\begin{lstlisting}
import java.io.File;

public class Files {
    /**
     * COMMENT:
     * Get all file names in the directory
     *
     * TEST:
     * import java.io.File;
     * import java.util.List;
     * import java.util.ArrayList;
     * __solution__
     * name_list = dfs("test_dir");
     * return(
     * name_list.contains("bar.txt") &&
     * name_list.contains("test2.txt"));
     */
    public List dfs(String dirname) {
        List result = new ArrayList();

        File dir_file = new File(??);
        ??

        return result;
    }
}
\end{lstlisting}
  \newpage
\item[Hello World GUI Draft] \hfill \\
\begin{lstlisting}
public class HelloWorld {

  /**
   * COMMENT:
   * Creating a program with buttons and counters
   *
   * TEST:
   * API_cons("hello-world-gui-test.java");
   * __solution__
   * source("src/main/resources/benchmarks/hello/test.java");
   * run();
   * return(
   *   has_frame &&
   *   has_label &&
   *   has_button &&
   *   has_listener &&
   *   added_label &&
   *   added_button &&
   *   set_close_op &&
   *   is_visible &&
   *   is_packed);
   */
  public void run() {
      JJFrame frame;
      JJLabel label;

      //Create and set up the window.
      frame = new JJFrame(??);
      ??

      int counter = 0;
      JJButton btn = new JJButton("increase");
      btn.addActionListener(new JActionListener() {
          public void actionPerformed(JActionEvent e) {
              counter += 1;
              label.setText(Integer.toString(counter));
          }
      });
      frame.getContentPane().add(btn);

      ??
  }


}
\end{lstlisting}
  \newpage
\item[HTTP Server Draft] \hfill \\
\begin{lstlisting}
import java.io.BufferedReader;
import java.io.IOException;
import java.nio.file.Files;

import com.sun.net.httpserver.HttpExchange;
import com.sun.net.httpserver.HttpHandler;
import com.sun.net.httpserver.HttpServer;

public class HTTP {

    /**
     * COMMENT:
     * Create an HTTP server which serves the content of a local file
     *
     * TEST:
     * import com.sun.net.httpserver.HttpExchange;
     * import com.sun.net.httpserver.HttpHandler;
     * import com.sun.net.httpserver.HttpServer;
     * import java.io.OutputStream;
     * __solution__
     * Random rand = new Random(System.currentTimeMillis());
     * int p = rand.nextInt(20000) + 35000;
     * HttpServer server = http("http_test.txt", p, "test");
     * test_url(new URL("http://localhost:" + Integer.toString(p) + "/test"));
     * server.stop(0);
     */
    public HttpServer http(String filename, int port, String url) {
        String content;

        /**
         * COMMENT:
         * Create an HTTP server which serves the content of a local file
         *
         * TEST:
         * String content;
         * String filename = "http_test.txt";
         * __solution__
         * return(content.equals("print(true);"));
         */
        ??

        HttpServer server;
        HttpHandler handler = new HttpHandler() {
            public void handle(HttpExchange he) throws IOException {
   			  he.sendResponseHeaders(??, content.length());
              OutputStream os = he.getResponseBody();
              os.write(content.getBytes());
              os.close();
            }
        };

        // set up the HTTP server
        ??
        return server;
    }
}
\end{lstlisting}
  \newpage
\item[LCS Draft] \hfill \\
\begin{lstlisting}
public class LCS {
    /**
     * COMMENT:
     * Longest common subsequence algorithm
     *
     * TEST:
     __solution__
     int NIL = -1;
     public boolean array_equal(int[] a1, int[] a2) {
       for(int i = 0; i < a1.length; ++i) {
         if(a1[i] != a2[i]) { return false; }
       }
       return true;
     }

     public int max(int a, int b) {
       return a > b ? a : b;
     }

     int[] test_array1 = {2, 6, 8, 10};
     int[] test_array2 = {1, 6, 8, 10};
     int[] test_ans = {6, 8};
     int[] result = new int[2];
     for(int j = 0; j < 2; ++j) {
       result[j] = -1;
     }

     lcs(test_array1, test_array2, 4, 4, result);
     if(!array_equal(result, test_ans)) {
       return(false);
     } else {
       return(true);
     }
     */
    void lcs(int[] X, int[] Y, int m, int n, int[] result) {
        int[][] L = new int[15][15];
        int i;
        int j;
        int index;
        int s;
        // build the lcs table
        ??

        index = L[m][n];
        s = index;
        i = m;
        j = n;
        while((i > 0) && (j > 0)) {
            if(X[(i - 1)] == Y[(j - 1)]) {
                result[(index - 1)] = X[(i - 1)];
                i --;
                j --;
                index --;
            } else {
                if(L[(i - 1)][j] > L[i][(j - 1)]) {
                    i --;
                } else {
                    j --;
                }
            }
        }
    }
}
\end{lstlisting}
  \newpage
\item[Matrix Multiplication Draft] \hfill \\
\begin{lstlisting}
public class MatMul {
    /**
     * COMMENT:
     * Matrix multiplication
     *
     * TEST:
     import java.util.Random;
     __solution__
     int[][] m1 = new int[5][5];
     int[][] m2 = new int[5][5];
     int[][] m3 = new int[5][5];
     int[][] m4 = new int[5][5];
     int[][] result = new int[5][5];
     int[][] ans = new int[5][5];

     public void my_mul(int[][] A, int[][] B, int[][] C, int n) {
       // matrix multiplication
       int i, j, k;
       int s;
       for (i=0; i<n; i++) {
         for (j=0; j<n; j++) {
           s=0;
           for (k=0; k<n; k++)
             s=s+A[i][k]*B[k][j];
           C[i][j]=s;
         }
       }
     }

     public boolean test_mat(int[][] m1, int[][] m2, int n) {
	   for(int i = 0; i < n; ++i) {
           for(int j = 0; j < n; ++j) {
               if(m1[i][j] != m2[i][j]) {
                   return false;
               }
           }
       }
	   return true;
     }

     for(int i = 0; i < 5; ++i){
       for(int j = 0; j < 5; ++j) {
         m1[i][j] = i + j; m2[i][j] = i + j;
         m3[i][j] = i + j; m4[i][j] = i + j;
       }
     }

     matmul(m1, m2, result, 5);
     my_mul(m3, m4, ans, 5);

     if(!test_mat(result, ans, k + 1)) {
       return(false);
     } else {
       return(true);
     }
     */
    void matmul(int[][] A, int[][] B, int[][] C, int n) {
			int i, j, k;
			int s;
			for (i=0; i<n; i++) {
				for (j=0; j<n; j++) {
                    // initialize the variable
					s=??;

                    // do the multiplication
					??
					C[i][j]=s;
				}
			}
    }
}
\end{lstlisting}
  \newpage
\item[Prim's Distance Update Draft] \hfill \\
\begin{lstlisting}
public class Prims {
    /**
     * COMMENT:
     * Prims algorithm
     *
     * TEST:
     int NIL = 100000;
     __solution__

     public boolean array_equal(int[] a1, int[] a2) {
       if(a1.length != a2.length) { return false; }
       for(int i = 0; i < a1.length; ++i) {
         if(a1[i] != a2[i]) { return false; }
       }
       return true;
     }

     int[][] test_graph = {{0, 2, 5, NIL, 8},
                           {NIL, 0, NIL, NIL, 1},
                           {NIL, NIL, 0, 4, NIL},
                           {NIL, NIL, NIL, 0, NIL},
                           {NIL, NIL, NIL, 3, 0}};

     int[][] test_loc = {{5, 2}, {5, 3}, {5, 4}};

     int[] ans = {2, 3, 4};
     int[][] result_cluster = {{0, 0}, {0, 0, 0}, {0, 0, 0, 0}}

     int[][] ans_cluster = {{0, 1}, {0, 1, 4}, {0, 1, 4, 3}};

     for(int i = 0; i < 3; ++i) {
         int test_size = test_loc[i][1];
         int result = prims(test_graph, result_cluster[i], test_loc[i][0], test_size);

         if(!array_equal(ans_cluster[i], result_cluster[i])) {
             return(false);
             exit();
         }
     }
     return(true);
     */
    public int prims(int[][] graph, int[] cluster, int nsize, int csize) {
        int[] key = new int[5];   // Key values used to pick minimum weight edge in cut
        boolean[] used = new boolean[5];
        int cluster_size = 0;
        int u;
        int min_dist;
        int count;

        // Initialize all keys as INFINITE
        for (int i = 0; i < nsize; i++)
          key[i] = ??;

        // Always include first 1st vertex in MST.
        key[0] = 0;     // Make key 0 so that this vertex is picked as first vertex

        // The MST will have V vertices
        for (count = 0; count < nsize; count++) {
            // Pick thd minimum key vertex from the set of vertices
            // not yet included in MST
            u = -1;

            int dist = 2147483647;
            for((min_dist = 0); (min_dist < nsize); (min_dist ++))
            {
                if((!used[min_dist]) && (dist > key[min_dist]))
                {
                    u = min_dist;
                    dist = key[min_dist];
                }
            }

            if(u == -1) { break; }

            // Add the picked vertex to the cluster
            used[u] = true;
            cluster[cluster_size++] = u;
            if(cluster_size == csize) {
                return cluster_size;
            }

            // update the node distances
            /**
             * COMMENT:
             * Prims update
             * 
             * TEST:
             * int u = 0;
             * int nsize = 4;
             * boolean[] used = {true, false, false, true};
             * int[] key = {10, 10, 10, 10};
             * int[][] graph = {{0, 100, 1, 0}};
             * __solution__
             * return(key[0] == 10 && key[1] == 10 && key[2] == 1 && key[3] == 10);
             */
            ??
        }

        return cluster_size;
    }
}
\end{lstlisting}
  \newpage
\item[Quick Sort Draft] \hfill \\
\begin{lstlisting}
public class QSort {
    /**
     * COMMENT:
     * quick sort partition
     *
     * TEST:
     * __solution__
     int NIL = 0;
     public boolean array_equal(int[] a1, int[] a2) {
       if(a1.length != a2.length) { return false; }
       for(int i = 0; i < a1.length; ++i) {
         if(a1[i] != a2[i]) { return false; }
       }
       return true;
     }

     public void swap(int[] array, int p1, int p2) {
       int tmp = array[p1];
       array[p1] = array[p2];
       array[p2] = tmp;
     }

     int[] test_array = {6, 2, 10, 8};
     int[] test_ans = {2, 6, 8, 10};
     int[] result = new int[4];

     for(int i = 0; i < 4; ++j) {
       result[i] = test_array[i];
     }
     qsort(result, 0, 3);
     if(!array_equal(test_ans, result)) {
       return(false);
     } else {
       return(true);
     }
     */
    void qsort(int[] a, int l, int r) {
        int j;
        int m = l;
        int i = ??;
        ??
        qsort(a, l, m - 1);
        qsort(a, m + 1, r);
    }
}
\end{lstlisting}
  \newpage
\item[Sieve Prime Draft] \hfill \\
\begin{lstlisting}
public class Sieve {

    /**
     * COMMENT:
     * Use Sieve of Eratosthenes algorithm to test primality
     *
     * TEST:
     * __solution__
     * return(
     * sieve(1) == false &&
     * sieve(2) == true &&
     * sieve(3) == true &&
     * sieve(4) == false &&
     * sieve(9) == false &&
     * sieve(17) == true &&
     * sieve(27) == false
     * );
     */
    boolean sieve(int n) {
        boolean[] primes = new boolean[100];
        for(int i = ??; i <= n; i++) {
            primes[i] = ??;
        }
        ??

        return primes[n];
    }
}
\end{lstlisting}
  \newpage
\end{description}

\end{document}